\documentclass[reprint,superscriptaddress,amsmath,amssymb,aps]{revtex4-2}

\usepackage{dcolumn}
\usepackage{bm}
\usepackage{graphicx}
\usepackage{epsfig}
\usepackage{amssymb}
\usepackage{amsmath, amscd}  
\newcommand{\R}{\mathbb{R}}

\newcommand{\C}{\mathbb{C}}
\usepackage{lipsum}
\newenvironment{itquote}
  {\begin{quote}\itshape}
  {\end{quote}\ignorespacesafterend}

\begin{document}

\title{Dynamics of a particle in the double-slit experiment with measurement}

\author{Alexey A. Kryukov \\  \footnotesize{Department of Mathematics \& Natural Sciences, University of Wisconsin-Milwaukee, USA \\ kryukov@uwm.edu}}

\begin{abstract}
Spontaneous collapse models use non-linear stochastic modifications of the Schr{\"o}dinger equation to suppress superpositions of eigenstates of the measured observable and drive the state to an eigenstate. It was recently demonstrated that the Born rule for
transition probabilities can be modeled using the linear Schr{\"o}dinger equation with a Hamiltonian represented by a random matrix from the Gaussian unitary ensemble. The matrices representing the Hamiltonian at different time points throughout the observation period are assumed to be independent. Instead of suppressing superpositions, such Schr{\"o}dinger evolution makes the state perform an isotropic random walk on the projective space of states. 
The relative frequency of reaching different eigenstates of an arbitrary observable in the random walk is shown to satisfy the Born rule.
Here, we apply this methodology to investigate the behavior of a particle in the context of the double-slit experiment with measurement. Our analysis shows that, in this basic case, the evolution of the particle's state can be effectively captured through a random walk on a two-dimensional submanifold of the state space. 
This random walk reproduces the Born rule for the probability of finding the particle near the slits, conditioned on its arrival at one of them. 
To ensure that this condition is satisfied, we introduce a drift term representing a change in the variance of the position observable for the state. It is argued that the drift term accounts for the energy transfer and trapping incurred during the particle's interaction with the detector. 
A drift-free model, based on equivalence classes of states indistinguishable by the detector, is also considered.
The resulting random walk, with or without drift, serves as a suitable model for describing the transition from the initial state to an eigenstate of the measured observable in the experiment, offering new insights into its potential underlying mechanisms. 
\end{abstract}

\maketitle

\section{Popular summary}

The superposition principle of Schr{\"o}dinger mechanics is foreign to Newtonian mechanics. Macroscopic objects are not observed in two different places, and the cat is not alive and dead simultaneously. However, such states are commonplace in the microworld. The debate on reconciling quantum and classical physics has continued since the early days of quantum mechanics. 
Here, we propose a new approach to the problem that allows us to deduce the Newtonian behavior of macroscopic particles and establish a connection between quantum and classical measurements, starting from the Schr{\"o}dinger equation with a random Hamiltonian.

In the proposed model, Newtonian motion emerges from Schr{\"o}dinger evolution by constraining the state of the particle to a certain part of the space of all its possible states. Mathematically, this part includes the usual 3-dimensional space of possible positions of the particle. On this 3-space, the Born rule, which gives the probability of finding the particle at a certain point in quantum theory, is equivalent to the classical probability law. Conversely, the classical probability law on the 3-space implies the Born rule on the space of states. Moreover, the Schr{\"o}dinger evolution that accounts for random fluctuations of the state of the measured particle becomes the Brownian motion of the particle, modeling the process of measurement in classical physics. In this setting, the superposition principle does not create a problem because superpositions of states follow the same evolution and satisfy the same Born rule. 

It is important to emphasize at the outset that the approach presented here is fully consistent with the well-established frameworks of quantum measurement and quantum decoherence theories. Rather than conflicting with these theories, it proposes a novel mechanism for state reduction, offering alternative methods for analyzing and describing quantum measurement and the quantum-to-classical transition.

We provide details of the state evolution in the double-slit experiment, considering both cases where the particle's position by the slits is measured and where it is not. The wave and corpuscular properties of the particle in the model are clarified, and their transition during measurement is explained.
It is shown that extending the classical 3-space of everyday experience to the space of states offers a promising route toward unifying classical and quantum perspectives and provides an alternative interpretation of the famous double-slit experiment.

\section{Prerequisites}

The Newtonian dynamics of a mechanical system can be identified with Schr{\"o}dinger dynamics under a constraint. The latter bears resemblance to the dynamics of a constrained classical system, like a bead on a wire. However, given that Schr{\"o}dinger dynamics is the dynamics of a quantum state, the constraint is now applied directly to the system's state.
For instance, consider a single-particle system in $\mathbb{R}^3$ described by the Hamiltonian ${\widehat h}=-\frac{\hbar^{2}}{2m}\Delta+{\widehat V}({\bf x},t)$.
The variation of the functional
\begin{equation}
\label{SS}
S[\varphi]=\int \overline{\varphi}({\bf x}, t) \left[i\hbar \frac{\partial}{\partial t}-{\widehat h}\right] \varphi({\bf x}, t) d^3 {\bf x} dt
\end{equation}
yields the Schr{\"o}dinger equation for the state function $\varphi$ of coordinates and time. Let $M^{\sigma}_{3,3}$ be the submanifold of the space of states $\mathbb{CP}^{L_2}$ of the particle formed by the states
\begin{equation}
\label{phii}
\varphi({\bf x})=g_{{\bf a}, \sigma}({\bf x})e^{i{\bf p}{\bf x}/\hbar}.
\end{equation}
Here
\begin{equation}
\label{g}
g_{{\bf a}, \sigma}=\left(\frac{1}{2\pi\sigma^{2}}\right)^{3/4}e^{-\frac{({\bf x}-{\bf a})^{2}}{4\sigma^{2}}}
\end{equation}
is the Gaussian function of a sufficiently small variance $\sigma^2$ centered at a point ${\bf a}$ in the Euclidean space $\R^3$, and ${\bf p}$ is a vector in $\R^3$.
For the states $\varphi$ constrained to the manifold $M^{\sigma}_{3,3}$, the functional (\ref{SS}) reduces to the classical action for the particle
\begin{equation}
\label{SSS}
S=\int \left[{\bf p}\frac{d {\bf a}}{dt}-h({\bf p},{\bf a},t)\right]dt.
\end{equation}
Here $h({\bf p},{\bf a},t)=\frac{{\bf p}^2}{2m}+V({\bf a},t)$ is the Hamiltonian function for the system. It follows that the variation of the functional (\ref{SS}) subject to the constraint that the state function $\varphi$ is in $M^{\sigma}_{3,3}$ yields Newtonian equations of motion. 
The appropriate value of the parameter $\sigma$ in the obtained relationship is dictated by the resolution of the position measuring instruments used for the particle to which this relationship is applied. 

Furthermore, in the derivation of the classical action (\ref{SSS}) for the particle, the Gaussian states $g_{{\bf a}, \sigma}$ in (\ref{phii}) can be replaced by arbitrary states of the form  $r_{{\bf a},\sigma}({\bf x})=\sigma^{-\frac{3}{2}}r(\frac{{\bf x}-{\bf a}}{\sigma})$. Here  $r \in L_2(\mathbb{R}^3)$ is any real-valued, twice-differentiable, unit-normalized function with finite variance (assumed to be $1$). This substitution yields the same result because the sequence $r^2_{{\bf a},\sigma}$, as $\sigma$ decreases, for any such function converges to the delta function $\delta^3_{\bf a}$ \cite{delta}.
Namely, through the direct substitution of the functions $\varphi({\bf x})=r_{{\bf a},\sigma}({\bf x})e^{i{\bf p}{\bf x}/\hbar}$ with ${\bf a}={\bf a}(t)$ and ${\bf p}={\bf p}(t)$ into (\ref{SS}) and noting that $\int r_{{\bf a},\sigma}({\bf x}) \nabla_{\bf a} r_{{\bf a},\sigma}({\bf x})d{\bf x}=0$ due to the normalization of $r_{{\bf a},\sigma}$, one deduces that, under these conditions, the convergence of $r^2_{{\bf a},\sigma}$ to the delta function is sufficient for deriving (\ref{SSS}). 
Consequently, the manifold $M^{\sigma}_{3,3}$ can be defined in terms of the functions $r_{{\bf a},\sigma}({\bf x})e^{i{\bf p}{\bf x}/\hbar}$, or in terms of equivalence classes of sufficiently narrow such functions.

It is well known that the Schr{\"o}dinger evolution of coherent states provides an approximation of the Newtonian dynamics of a particle. For a quadratic potential, this approximation remains valid at all times. Furthermore, maintaining the Gaussian form of the state allows for a good semiclassical approximation of Schr{\"o}dinger evolution \cite{Gauss}. However, the idea that the coherent form of the states (\ref{phii}) is solely responsible for the Newtonian behavior of particles confined to the manifold $M^{\sigma}_{3,3}$ has been refuted in the previous paragraph.
At the same time, the result is not unexpected, as it aligns with the predictions of the Ehrenfest theorem for sufficiently narrow wave packets, provided this narrow form is preserved throughout the evolution.

By confining the state to the manifold $M^{\sigma}_{3,3}$, defined in terms of arbitrary sufficiently narrow functions $r_{{\bf a},\sigma}({\bf x})e^{i{\bf p}{\bf x}/\hbar}$, we establish the relationship between Schr{\"o}dinger and Newtonian evolution, a connection that persists for all potentials and time intervals. In essence, by identifying the source of the constraint on the state of macroscopic bodies, a goal we aim to pursue in this paper, we can unify the dynamics of both microscopic and macroscopic entities. Note that, for the sake of simplicity, we will continue using functions $g_{{\bf a},\sigma}$ throughout most of the paper. The convergence of the functions $r^2_{{\bf a},\sigma}$ to the delta function indicates that integral expressions involving these functions should approach the same limit as the specific case where $r_{{\bf a},\sigma}=g_{{\bf a},\sigma}$.
Whenever a different function choice within the equivalence class becomes important, we will address the matter.

 The Fubini-Study metric on $\mathbb{CP}^{L_2}$ provides a Riemannian metric on $M^{\sigma}_{3,3}$. 
 With an appropriate choice of units, the map $\Omega: ({\bf a}, {\bf p}) \longrightarrow g_{{\bf a}, \sigma}e^{i{\bf p}{\bf x}/\hbar}$ serves as an isometry between the Euclidean space $\R^3 \times \R^3$ and the Riemannian manifold $M^{\sigma}_{3,3}$. 
If desired, a linear structure on $M^{\sigma}_{3,3}$ can be induced by $\Omega$ from the one on $\R^3 \times \R^3$. The restricted map $\omega: {\bf a} \longrightarrow g_{{\bf a}, \sigma}$ acts as an isometry between the Euclidean space $\R^3$ and the Riemannian submanifold $M^{\sigma}_3$ of $\mathbb{CP}^{L_2}$ formed by the states $g_{{\bf a}, \sigma}$ \cite{KryukovMath, KryukovPhysLett, Kryukov2020}. 
This remains true for the functions $r_{{\bf a},\sigma}$ in place of $g_{{\bf a}, \sigma}$.
The relationship between action functionals (\ref{SS}) and (\ref{SSS}) enables us to identify classical particles, i.e., particles that satisfy Newtonian dynamics, with quantum systems whose state is constrained to the manifold $M^{\sigma}_{3,3}$ with an appropriate value of $\sigma$. The map $\Omega$ identifies the Euclidean phase space $\R^3 \times \R^3$ of positions and momenta $({\bf a}, {\bf p})$ of a classical particle with the manifold $M^{\sigma}_{3,3}$ of states $\varphi$ in (\ref{phii}).
Imposing the constraint amounts to making the components of the velocity of state $\frac{d \varphi}{dt}=-\frac{i}{\hbar}{\widehat h}\varphi$ orthogonal to the manifold $M^{\sigma}_{3,3}$ vanish. The components tangent to $M^{\sigma}_{3,3}$ correspond to the Newtonian velocity and acceleration of the particle. Commutators of observables become Poisson brackets, transforming the Schr{\"o}dinger dynamics of the constrained state into the Newtonian dynamics of the particle \cite{KryukovNew}.

The embedding of classical space and classical phase space into the space of states resulted in a relationship between Schr{\"o}dinger and Newtonian dynamics. This relationship enabled us to identify classical particles with quantum systems whose state is constrained to the manifold $M^{\sigma}_{3,3}$.  The value of the parameter $\sigma$ is determined by the resolution of the measuring instruments used. Let us show that the embedding and identification also lie at the core of the relationship between the normal probability distribution, typical for position measurements of a particle in $\R^3$, and the Born rule governing the probability of transitions between states. 

Let $\delta$ denote the diameter of a small region $\mathit W$ that contains the point ${\bf b}$ in $\R^3$. Suppose the measured position of a classical particle is normally distributed and centered at ${\bf a}$, so that the probability of finding the particle in $\mathit W$ is approximately the product of the normal probability density function and the volume of the region. The isomorphism $\omega$ identifies $\R^3$ with the manifold $M^{\sigma}_3$. The classical particle at ${\bf a}$ is represented by the state $g_{{\bf a}, \sigma}$.
Let us now show that applying the Born rule to the state $g_{{\bf a}, \sigma}$ yields the normal distribution for the classical particle's position on $\R^3$. Conversely, assuming a normal probability distribution for the particle's position in $\R^3$, and considering the corresponding probability of finding the particle within region $\mathit W$, one recovers the Born rule for transitions between the associated states. If the transition probabilities depend solely on the Fubini-Study distance between states, this result extends to all transitions within the state space.

The first part of the claim follows from the fact that for states $g_{{\bf a}, \sigma}$ in $M^{\sigma}_{3}$ or states $\varphi$ in $M^{\sigma}_{3,3}$, the probability density $|g_{{\bf a}, \sigma}|^2$ in the Born rule is also the normal probability density function on $\R^3$. Due to the identification of classical particles with quantum systems whose state is constrained to the manifold $M^{\sigma}_{3,3}$, we conclude that the Born rule on $\mathbb{CP}^{L_2}$ implies the normal probability distribution of the particle's position on $\R^3=M^{\sigma}_{3}$.

To prove the second part, let $\rho(g_{{\bf a}, \sigma}, g_{{\bf b}, \delta})$ denote the Fubini-Study distance between the Gaussian states $g_{{\bf a}, \sigma}$ and $g_{{\bf b}, \delta}$. 
Through direct integration, we obtain:
\begin{equation}
\label{del}
\left(\frac{2\sigma \delta}{\sigma^2+\delta^2}\right)^3 e^{-\frac{({\bf a}-{\bf b})^2}{2(\sigma^2+\delta^2)}}=\cos^{2}\rho(g_{{\bf a}, \sigma}, g_{{\bf b}, \delta}).
\end{equation}
If $\delta=\sigma$, this equation establishes a relationship between the distances between states $g_{{\bf a}, \sigma}$ and $g_{{\bf b}, \sigma}$ in the Fubini-Study metric on $\mathbb{CP}^{L_2}$ and points ${\bf a}$ and ${\bf b}$ in the Euclidean metric on $\R^3$:
\begin{equation}
\label{mainO}
e^{-\frac{({\bf a}-{\bf b})^{2}}{4\sigma^{2}}}=\cos^{2}\rho(g_{{\bf a}, \sigma}, g_{{\bf b}, \sigma}).
\end{equation}
Note that the distance $\rho(g_{{\bf a}, \sigma}, g_{{\bf b}, \sigma})$ is measured along a geodesic in the full state space, whereas the Euclidean distance $|{\bf a} - {\bf b}|$ corresponds to the distance between the same states measured along a geodesic within the submanifold $M^{\sigma}_3$.
The distance between the states $\varphi({\bf x})=g_{\bf a}({\bf x})e^{i{\bf p}{\bf x}/\hbar}$ and $\psi({\bf x})=g_{\bf b}({\bf x})e^{i{\bf q}{\bf x}/\hbar}$, measured using the Fubini-Study metric on $\mathbb{CP}^{L_2}$, is related to the Euclidean distance between the corresponding points in the classical phase space $\R^3 \times \R^3$ by a similar formula:
\begin{equation}
\label{mainOO}
e^{-\frac{({\bf a}-{\bf b})^{2}}{4\sigma^{2}}-\frac{({\bf p}-{\bf q})^{2}}{\hbar^2/\sigma^{2}}}=\cos^{2}\rho(\varphi, \psi).
\end{equation}

On another hand, when $\delta$ in (\ref{del}) approaches $0$, the left side of (\ref{del}) yields the normal probability density function times the volume element $(8\pi)^\frac{3}{2}\delta^3$. 
Due to the identification of classical particles with quantum systems whose states are constrained to the manifold $M^{\sigma}_{3,3}$, the result can be interpreted as the probability of finding the particle in the region ${\mathit W}$, assuming that the distribution of the measured position is normal and centered at ${\bf a}$.  For instance, the paper models the measurement of a classical particle's position in $\R^3$ as a random walk from its initial value ${\bf a}$ during the observation period, resulting in a normal distribution of the position. 
The probability on the right side of (\ref{del}) is the probability of transition between the corresponding initial and end states, calculated by the Born rule. 

The Fubini-Study distance between states $g_{{\bf a}, \sigma}$ and  $g_{{\bf b}, \delta}$ with ${\bf a}$ and ${\bf b}$ in $\R^3$ takes on all possible values  in $\mathbb{CP}^{L_2}$, ranging from $0$ to $\pi/2$. Let then $\varphi$ and $\psi$ be any two states in $\mathbb{CP}^{L_2}$, and 
let $g_{{\bf a}, \sigma}$ and $g_{{\bf b}, \delta}$ represent two states at the same Fubini-Study distance as the distance between $\varphi$ and $\psi$. 
 By assumption, the probability of transition between two states depends only on the distance between them.
 The probability $P({\varphi, \psi})$ of transition between $\varphi$ and $\psi$ is then given by:
\begin{equation}
\label{distt}
P({\varphi, \psi})=P({g_{{\bf a}, \sigma}, g_{{\bf b}, \delta}})=\cos^{2}\rho(g_{{\bf a}, \sigma}, g_{{\bf b}, \delta})= 
  \cos^{2}\rho(\varphi, \psi).
\end{equation}
So, $P({\varphi, \psi})=\cos^{2}\rho(\varphi, \psi)$, which is the Born rule. We conclude that under these conditions, the normal probability distribution on $\R^3$ implies the Born rule on the space of states.
Note that while the proof relies on the Gaussian form of the functions $g_{{\bf a}, \sigma}$, the result is general and will be shown to originate from the connection between the dynamics of the state in the full state space and the submanifold $M^{\sigma}_{3,3}$.


The correspondence established between classical and quantum systems, and between normal probability distribution and the Born rule was leveraged in \cite{KryukovNew} to put measurements performed on classical and quantum systems on an equal footing.
To achieve this, the following proposition, based on Wigner's work \cite{Wigner} and the Bohigas-Giannoni-Schmit conjecture \cite{BGS}, and further expounded upon in \cite{KryukovNew}, was introduced:
\begin{itquote}{\bf{(RM)}}
The dynamics of a particle's state under position measurement can be modeled as a random walk in the space of states. 
In the absence of drift, the steps of this random walk satisfy the Schr{\"o}dinger equation, where the Hamiltonian at any given time is represented by a random matrix from the Gaussian Unitary Ensemble (GUE). The matrices representing the Hamiltonian at different times are statistically independent.
\end{itquote}
Here, the abbreviation RM in {\bf (RM)} refers to ``random matrices". Physically, the Hamiltonian in {\bf (RM)} may arise from a highly complex interaction between the measured particle and the measuring device or environment, modeled as a complicated sum of one-particle Hamiltonians with interaction terms. 
This is reminiscent of Wigner's model for the Hamiltonian of a heavy nucleus.

The Gaussian unitary ensemble consists of Hermitian matrices whose entries on and above the diagonal are independent random variables. The entries above the diagonal are identically distributed normal complex random variables, whose real and imaginary parts have mean $0$ and variance $d^2$. The diagonal entries are real normal random variables with mean $0$ and variance $2d^2$. Such matrices can be expressed in the form $\frac{1}{\sqrt{2}}(A+A^{\ast})$, where $A$ is a square matrix whose entries are independent, identically distributed complex normal random variables, and $A^{\ast}$ is the Hermitian conjugate of $A$. The central characteristic of the Gaussian unitary ensemble is that the probability density function $P$ on matrices ${\widehat h}$ within the ensemble remains invariant under unitary transformations: $P(U^{\ast}{\widehat h}U)=P({\widehat h})$  \cite{Fyodorov}.

A small step in the walk of state driven by the Hamiltonian in {\bf (RM)} is a random vector in the tangent space to the space of states $\mathbb{CP}^{L_2}$. 
As demonstrated in \cite{KryukovNew}, the distribution of steps in the walk is normal, homogeneous, and isotropic. In particular, the orthogonal components of a step at any point are independent identically distributed normal random variables. From these properties, it follows that the probability of transition between two states connected by the walk may depend solely on the Fubini-Study distance between them. 
Under the condition that the steps of the walk occur on  $M^{\sigma}_3$, the probability of transition is determined by the normal probability density function. In this case, the random walk of the state approximates Brownian motion on $\R^3$, making it suitable for modeling classical measurement. Since the probability of transition $P(\varphi, \psi)$ between two states depends solely on the distance between them, and because the probability density function for the states $\varphi$ and $\psi$ in $M^{\sigma}_{3}$ is normal, we conclude, based on the derivation culminating in (\ref{distt}), that $P(\varphi, \psi)$ is governed by the Born rule \cite{KryukovNew}.
Consequently, both the normal probability distribution valid for classical measurements and the Born rule for the probability of transition between general quantum states arise from the Schr{\"o}dinger evolution with a Hamiltonian satisfying {\bf (RM)}. 

Because Brownian motion is governed by the diffusion equation, the dynamic underpinning of the Born rule and the normal probability distribution in the model can be expressed as follows: the Schr{\"o}dinger equation with the Hamiltonian in {\bf (RM)} reduces to the diffusion equation on $\R^3$. This assertion can be explicitly illustrated  by introducing the density of states functional on the state space. This approach also facilitates the derivation of the connection between the Born rule and the probability distribution on $\R^3$ in the presence of boundary conditions, leading to a non-normal distribution. We will briefly outline this method closely following \cite{Kryukov2020}. A more straightforward derivation of this result is provided in \cite{KryukovNew}.

In non-relativistic quantum mechanics, particles and their corresponding states in a single-particle Hilbert space cannot vanish or be created. The unitary nature of evolution dictates that states can only traverse the unit sphere in the space of states $L_{2}(\R^3)$. 
The normalized states resulting from a measurement also lie on the sphere.
Let us introduce the density of states functional $R_{t} [\varphi;\psi]$. Here, we start with an ensemble of particles whose initial state lies in a vicinity of the state $\psi$ in $\mathbb{CP}^{L_2}$. The functional $R_{t} [\varphi;\psi]$ quantifies the number of states that, by time $t$, reside in a vicinity of a state $\varphi \in \mathbb{CP}^{L_2}$. It approximates the count of states in a small region surrounding $\varphi$ in $\mathbb{CP}^{L_2}$, normalized by the volume of the region. Notably, measuring devices occupy finite regions and possess finite resolutions, rendering the effective space of states finite-dimensional, thereby allowing the existence of Lebesgue measure on the space \cite{Kryukov2020}.

Under the isometric embedding $\omega: \R^3 \longrightarrow M^{\sigma}_{3} \subset \mathbb{CP}^{L_2}$, the state functions in $M^{\sigma}_{3}$ correspond to Newtonian particle positions in $\R^3$. 
As a result, the functional $R_{t} [g_{\bf a};\psi]$ can be considered as a functional on functions $\psi$ dependent on the position ${\bf a}$. We can then normalize it by the volume of the region in $\R^3$. The resulting density of state functional will be labeled as $\rho_{t}[{\bf a};\psi]$.
For the same reason, the density of states functional $\rho_{t}[{\bf a};\psi]$ for $\psi$ in $M^{\sigma}_{3}$ must correspond to the conventional particle density $\rho_{t}({\bf a}; {\bf b})$ for particles initially positioned at ${\bf b}$ in $\R^{3}$, where their position is measured a short time later (mean observation period). In other words, we must have $\rho_{t}({\bf a}; {\bf b})=\rho_{t}[{\bf a}; g_{{\bf b},\sigma}]$.

Let us apply the density of states functional to the dynamics of macroscopic and microscopic particles, with or without measurement.
If $\rho_{t}({\bf a};{\bf b})$ represents the density of an ensemble of macroscopic particles at a point ${\bf a} \in \R^3$ with an initial position near ${\bf b}$, and ${\bf j}_{t}({\bf a}; {\bf b})$ denotes the current density of particles at ${\bf a}$, then the conservation of the number of particles
implies the continuity equation:
\begin{equation}
\label{conti}
\frac{\partial \rho_{t}({\bf a};{\bf b})}{\partial t}+\nabla \cdot {\bf j}_{t}({\bf a};{\bf b})=0.
\end{equation}
We can assume that $\rho_{t}({\bf a};{\bf b})$ and ${\bf j}_{t}({\bf a};{\bf b})$ are normalized per one particle, i.e., the densities are divided by the number of particles. In this case, the particle density and the probability density can be identified.

The continuity equation resulting from Schr{\"o}dinger dynamics with Hamiltonian ${\widehat h}=-\frac{\hbar^{2}}{2m}\Delta+{\widehat V}({\bf x},t)$ matches equation (\ref{conti}) with the substitutions:
\begin{equation}
\label{contiS}
\rho_{t}=|\psi |^2, \quad {\rm and} \quad {\bf j}_{t}=\frac{i\hbar}{2m}(\psi \nabla {\overline \psi}-{\overline \psi}\nabla \psi).
\end{equation}
For states $\psi \in M^{\sigma}_{3,3}$, we derive:
\begin{equation}
\label{psiB}
{\bf j}_{t}=\frac{\bf p}{m} |\psi|^2={\bf v} \rho_{t}.
\end{equation}
As we know, the Schr{\"o}dinger evolution of a particle's state constrained to the manifold $M^{\sigma}_{3,3}$ corresponds to Newtonian evolution. It is also known that, for states initially in $M^{\sigma}_{3,3}$, the imposition of this constraint is equivalent to eliminating the spreading component of the state's velocity (see (\ref{decomposition}) and \cite{KryukovNew}). As follows from (\ref{decomposition}), one way to achieve this is to consider the motion of particles of sufficiently large mass over short time intervals. In particular, for such particles, the continuity equation for the Schr{\"o}dinger evolution remains valid, but
must reduce to the continuity equation (\ref{conti}) for Newtonian motion. Because this result depends solely on the fact that the evolving state is constrained to $M^{\sigma}_{3,3}$, that is, the spreading component of the state's velocity vanishes, it holds regardless of the method used to enforce this constraint.

From (\ref{contiS}) and (\ref{psiB}), we see that for the continuity equation of the  conventional Schr{\"o}dinger evolution to reduce  to that of Newtonian motion, the density $\rho_t$ for states in $M^{\sigma}_{3,3}$ must represent the particle density $\rho_{t}({\bf a}; {\bf b})$.
This quantity represents the number of particles that originate from a neighborhood of ${\bf b}$ and, by the time of observation, reach a neighborhood of ${\bf a}$.
The relationship $\rho_{t}({\bf a}; {\bf b})=\rho_{t}[{\bf a}; g_{{\bf b},\sigma}]$ indicates that the general form of $\rho_{t}$ in (\ref{contiS}) can be identified with the density of states $\rho_{t}[{\bf a}; \psi]$. It represents the number of particles initially in a state near $\psi$ that, at the time of observation, reside in a state close to $g_{{\bf a},\sigma}$, indicating their proximity to the point ${\bf a}$.  
With this identification accepted, the flow of states on the state space describes both the flow of particles and the flow of probability on $\R^3$ as special cases. 
Note that the probability density in (\ref{contiS}) indicates the likelihood of the initial state $\psi$ reaching a state in $M^{\sigma}_{3}$, as opposed to any other state. Consequently, we examine the flow of states under the condition that, upon measurement, they reach the classical space submanifold $M^{\sigma}_{3}$.

The relation  $\rho_{t}[{\bf a}; \psi]=\rho_{t}$ together with (\ref{contiS}) imply that
\begin{equation}
\label{density}
\rho_{t}[{\bf a}; \psi]=|\psi({\bf a})|^2,
\end{equation}
which clarifies the association of $|\psi({\bf a})|^2$ with the probability density, a fundamental postulate in quantum theory. Specifically, the probability density of finding the system in a state, across an ensemble of states, is proportionate to the value of the density of states functional at that state, as given by (\ref{density}) for states $g_{{\bf a},\sigma}$ in $M^{\sigma}_{3}$. Hence, $|\psi({\bf a})|^2$ serves as the probability density of locating the particle near ${\bf a}$ because it represents the density of quantum states near the point $g_{{\bf a},\sigma}$ at the time of observation. As the number of states near $g_{{\bf a},\sigma}$ increases, the likelihood of observing the state near that point also increases.

Note that the flow of states during the measurement process should not be confused with the probability flow under conventional Schr{\"o}dinger evolution. 
Thus far, we have considered the flow of states governed by standard Schr{\"o}dinger evolution, where $t$ serves as the time parameter. 
Nevertheless, the continuity equation remains applicable during measurement governed by {\bf (RM)} as well, since the total probability of finding the particle is conserved.
From experience, we know that the time $\tau$ required for the initial state $\psi$ to transition to the measured state $g_{{\bf a},\sigma}$ is extremely short. This justifies the approximation $t+\tau \approx t$, which was implicitly used in the interpretation of equation (\ref{density}).

To prove the equality (\ref{density}) and explain the Gaussian form of $\rho_{t}({\bf a}; {\bf b})$ during position measurement in the model, let us analyze the dynamics of the flow of states generated by the Schr{\"o}dinger equation with the Hamiltonian in {\bf (RM)}.
This equation dictates how the density of states functional, initially concentrated at the point $\psi$, diffuses throughout the space of states while adhering to the Born rule. 
Expressed in integral form, the conservation of states in $\mathbb{CP}^{L_2}$ takes the following form:
\begin{equation}
\label{Fdiffusion}
R_{t+\tau}[\varphi;\psi]=\int R_{t}[\varphi+\eta;\psi]\gamma[\eta] D\eta,
\end{equation}
where $\gamma[\eta]$ represents the probability functional of the variation $\eta$ in the state $\varphi$, and the integration is over all variations $\eta$ such that $\varphi+\eta \in \mathbb{CP}^{L_2}$. Due to the homogeneity and isotropy of the distribution of steps in the random walk generated by the Hamiltonian in {\bf (RM)}, $\gamma[\eta]$ solely depends on the Fubini-Study distance between $\varphi+\eta$ and $\varphi$, and not the point $\varphi$ or the direction of $\eta$.

Let us demonstrate that when the particle's state is confined to $M^{\sigma}_{3}=\R^3$, this equation implies the conventional diffusion equation on $\R^3$. When (\ref{Fdiffusion}) is restricted to $M^{\sigma}_{3}$, we have $\psi=g_{{\bf b},\sigma}$ and $\eta=g_{{\bf a}+{\bf e}, \sigma}-g_{{\bf a},\sigma}$, where ${\bf e}$ represents a displacement vector in $\R^3$. As previously established, the functional $\rho_{t}[{\bf a};g_{{\bf b},\sigma}]$ is identified with the conventional density of particles in space $\rho_{t}({\bf a};{\bf b})$.
Substituting this into (\ref{Fdiffusion}), and replacing $\gamma[\eta]$ with the equivalent probability density function $\gamma({\bf e}) \equiv \gamma[g_{{\bf a}+{\bf e}, \sigma}-g_{{\bf a},\sigma}]$, we integrate over the space $\R^3$ of all possible vectors ${\bf e}$:
\begin{equation}
\label{Rdiffusion}
\rho_{t+\tau}({\bf a};{\bf b})=\int \rho_{t}({\bf a}+{\bf e};{\bf b})\gamma({\bf e}) d^3{\bf e}.
\end{equation}
Because $\gamma({\bf e})$ only depends on the norm of ${\bf e}$, this leads to the diffusion equation, in the same way as in Einstein's paper on Brownian motion \cite{Ein}:
\begin{equation}
\label{diffusionR}
\frac{\partial \rho_{t}({\bf a};{\bf b})}{\partial t}={\mathbb D}\Delta \rho_{t}({\bf a};{\bf b}),
\end{equation}
where ${\mathbb D}$ is the diffusion coefficient. 
The solution to (\ref{diffusionR}) for the particle initially at ${\bf b}$ yields the normal probability density function, which aligns with the choice of functions $g_{{\bf b},\sigma}$ in $M^{\sigma}_{3}$. Therefore, the relation (\ref{density}) dynamically follows from the resulting normal probability distribution on $\R^{3}$ and the derivation leading up to (\ref{distt}). 

The conjecture {\bf (RM)} specifies dynamics of microscopic particles under position measurement. When the dynamics is constrained to $M^{\sigma}_3$, it describes the behavior of macroscopic particles whose positions are being measured. Conversely, the constrained dynamics determines the probability distribution of the corresponding Hamiltonian entries in {\bf (RM)}, thereby uniquely specifying the entire ensemble. 
Later in the paper, it will be argued that the constraint to $M^{\sigma}_3$ may result from a drift in the random walk described in {\bf (RM)}. Alternatively, the apparent constraint may be related to the use of equivalence classes of states that are indistinguishable by the detector.
Thus, the conjecture {\bf (RM)}, with a possible drift term included, may be capable of addressing position measurements for both microscopic and macroscopic particles.

Let us emphasize once again that the diffusion equation and the Brownian motion it describes are used here to model the positional measurement of a classical particle. The time parameter in the probability density function for the position is identified with the mean observation period. While individual observation periods may vary, the mean period corresponds to the variance, $\sigma^2$, of the observed normal distribution of the particle's position. In turn, the variance depends on the resolution of the measuring device being used.
In the absence of imposed boundary conditions, solving the diffusion equation for a particle initially positioned at ${\bf b}$ yields a normal distribution. 
However, the diffusion equation remains an appropriate method for modeling the measurement of a classical particle's position, even in scenarios where boundary conditions like confinement in a box are applied. In such cases, the particle's inability to exist outside the box results in a probability distribution that differ from the normal distribution. This aligns with the solution to the diffusion equation, which satisfies the prescribed boundary conditions. 

The imposition of boundary conditions restricts the Hilbert space of possible particle states. For example, one might obtain the Hilbert space $L_2[a,b]$ of square-integrable functions on the interval $[a,b]$, rather than on the entire number line $\R$. In the scenario where the endpoints of the interval are absorbing, as in the infinite potential well, the admissible state functions belong to a subspace of functions that are zero at $a$ and $b$. 
These spaces are subspaces of the total Hilbert space of states, such as $L_2(\R)$, with no imposed boundaries. 
The random matrix representing the Hamiltonian remains in the Gaussian unitary ensemble, albeit now acting on a subspace of the original Hilbert space. The random walk in {\bf (RM)} takes place on this subspace. The distribution of steps of the walk continues to be homogeneous and isotropic in the new space. 

It follows that the probability of transition between states in the new space remains dependent solely on the Fubini-Study distance between them. Since in addition steps in orthogonal subspaces such as $L_{2}[a,b]$ and its orthogonal complement in $L_{2}(\R)$ are independent, the probability of transition between states will adhere to the same rule. 
For instance, relative frequencies of different measurement outcomes of the position in both $L_{2}[a, b]$ and $L_{2}(\R)$ spaces satisfy the Born rule. 
A comparison between the random walk of a Brownian particle on the plane $\R^2$ and that on the number line $\R$ serves to clarify this result. Namely, the probability of a Brownian particle on $\R^2$ reaching at a given time a rectangle of an arbitrary fixed height $\epsilon$ based on an interval $[c, d]$ along the $x$-axis, for all $c$ and $d$, is proportional to the probability of the particle constrained to $\R$ reaching the same interval. The relative frequencies of finding the particle in the rectangles and the intervals are the same.

The random walk defined in {\bf (RM)} on the projective space of the space $L_{2}(\R^3)$ constrained to the manifold $M^{\sigma}_{3}$ was shown to approximate a solution to the diffusion equation on $\R^3$. A similar result holds true for the walk on $L_{2}(\R)$ constrained to the one-dimensional submanifold $M^{\sigma}_{1}$ of $L_{2}(\R)$ formed by the corresponding Gaussian states $g_{c,\sigma}$. This correspondence between the walk in {\bf (RM)} and the diffusion equation resulted in the correspondence between the Born rule and the normal probability distribution law. When absorbing boundary conditions of interest here are imposed, the solution to the diffusion equation changes. 
We therefore need to identify a manifold of states in $L^2[a,b]$, denoted here by $M_{[a,b]}$, that represents the interval $[a,b]$ in $\R$, such that the random walk in {\bf (RM)} on the appropriate space of states constrained to $M_{[a,b]}$ yields the diffusion equation with the required initial and boundary conditions.

Analogous to the case of the manifolds  $M^{\sigma}_{3}=\R^3$ and $M^{\sigma}_{1}=\R$, the points $r_{c,\sigma}$ of $M_{[a,b]}$ will be represented by the square root of the solution to the diffusion equation with an initial point-source at $c$ and absorbing boundaries at $a$ and $b$. 
Since the variance $\sigma^2$ is small, the functions $r_{c,\sigma}$ representing points of $[a,b]$ that lie away from the boundaries 
 are well approximated by Gaussian functions $g_{c,\sigma}$. In fact, the distribution of small steps originating from such points, obtained by solving the diffusion equation with absorbing boundaries, is nearly normal. 
Moreover, for interior points, these Gaussian functions can be generated by translating a single such function. As before, the induced metric at these points is Euclidean. 
However, for points in small neighborhoods near the boundaries, the corresponding functions are ``squeezed" to satisfy the vanishing condition outside the interval.
The decreased value of $\sigma$ affects the induced metric near $a$ and $b$. As a result, the step sizes of the walk constrained to $M_{[a,b]}$ become vanishingly small near the boundaries. 
(For the inner product of squeezed Gaussian functions centered near the endpoints of $[a,b]$ to remain unchanged, the Euclidean distance between their centers must decrease. See equation (\ref{mainO}).)

Effectively, the real line $\R$ is compressed into the interval $(a,b)$, with points outside $[a,b]$ mapped into small neighborhoods near $a$ and $b$. The endpoints $a$ and $b$ thus correspond to $\pm\infty$ on $\R$ and serve as absorbing boundaries, i.e., points of no return. The random walk in {\bf (RM)}, when considered on $M_{[a,b]}$, becomes a walk with absorbing boundaries on $[a,b]$, which, in the proper limit, yields Brownian motion on $[a,b]$.
Meanwhile, the state driven by the Hamiltonian in {\bf (RM)} does not freeze but continues to spread within the state space. The boundary conditions are ingrained in the choice of the Hilbert space of states, while the properties of the walk in {\bf (RM)} remain unchanged.

We conclude that, with or without the considered boundary conditions, the random walk of states defined in {\bf (RM)} can be utilized to model the position measurement of both macroscopic and microscopic particles.
The Born rule emerges in two connected ways. When the classical space $\R^3$ is identified with the manifold $M^{\sigma}_{3}$, the Born rule emerges as the unique probability law that depends solely on the Fubini-Study distance between states and remains compatible with the normal distribution on $\R^3$. At a deeper level, the Born rule in the framework dynamically emerges from the homogeneity and isotropy of the probability distribution of steps in the random walk and its transition to a walk approximating a solution of the diffusion equation.

Until now, we have aimed to highlight the similarities between measurements in classical and quantum settings, putting them on equal footing. To illustrate the difference between these measurements, consider that the Brownian motion of a measured particle occurs in three-dimensional space, $\R^3$, where its position can be measured at any point. This is feasible because position-measuring devices can be evenly distributed throughout space. However, the same cannot be achieved in the space of states. Macroscopic position-measuring devices may only occupy a submanifold, such as $M^{\sigma}_{3}$ or a product of $n$-copies of it in the space of states. For the position of a microscopic particle to be defined and measured, its initial state must first traverse the classical space submanifold $M^{\sigma}_{3}$ in the space of states. In classical terms, an analogy to this scenario would be measuring the position of a Brownian particle in $\R^3$ using particle detectors arranged along a line not passing through the initial position of the particle. To ascertain its position in this case, the Brownian particle must first intersect the line. The probability of reaching a particular segment of the line reflects the information provided by the Born rule.

The model based on the Schr{\"o}dinger equation with a random Hamiltonian under discussion here requires comparison with spontaneous collapse models \cite{Bassi1, Bassi2}. These latter models were developed to address the process of measurement and the absence of macroscopic superpositions in our surroundings. They employ a non-linear stochastic modification of the Schr{\"o}dinger equation to drive the state towards an eigenstate of the measured observable, such as position, energy, momentum, or spin \cite{Adler, Benatti, Pearle}. Among these, only the models ensuring collapse in the position basis elucidate a definite position of macroscopic objects in space, as discussed herein. 
The principal models of this kind include GRW, CSL, and QMUPL models, reviewed in \cite{Bassi2}. 

All existing spontaneous collapse models inducing collapse in the position basis lead to the collapse of the state function in space. Collapse occurs more rapidly in larger systems. The stochastic nature of collapse in the models is frequently attributed to a random field in space that interacts with matter in a non-linear manner, leading to collapse. The noise associated with the field could be white, with all frequencies contributing equally to collapse, or Gaussian. The stochastic equation could also incorporate dissipative terms. The models may encompass only systems of distinguishable particles or systems of identical particles. The collapse dynamics in the models is equally applicable to all quantum processes, with or without measurement. Measurements in the models yield a single outcome, distributed according to the Born rule. In models lacking dissipative terms, the energy of the quantum system increases. Another common challenge of the spontaneous collapse models is their relativistic formulation, as collapse must be nearly instantaneous.

In contrast to existing collapse models, the Schr{\"o}dinger equation with the Hamiltonian in {\bf (RM)} is a linear stochastic equation. The evolution in the model does not disrupt superpositions but rather causes the state to meander through the entire space of states. 
It was shown that, for such an evolution, the probability of reaching a particular state in the classical space submanifold $M^{\sigma}_3$ conforms to the Born rule. 
The functions in $M^{\sigma}_3$ are approximate eigenstates of the position operator. 
By applying a unitary transformation on $L_2(\R^3)$, we alter the functional form of the position operator and replace the manifold $M^{\sigma}_3$ with the set of approximate eignenstates of the resulting operator. Under this mapping, the Hamiltonian in {\bf (RM)} retains its properties. Since distances are preserved, we conclude that the probability of reaching an eigenstate of the new operator remains consistent with the Born rule. 

 In particular, the probability of reaching a specific eigenstate of the momentum operator under the walk is in agreement with the Born rule.
Because the probability of reaching a state via the random walk in {\bf (RM)} over a given time interval depends only on the Fubini-Study distance between the initial and final states, the evolution described by {\bf (RM)} may be capable of correctly characterizing the probabilities of measurement outcomes for other observables on $L_2(\R^3)$.
An additional process may be required to drive the state toward the manifold of eigenstates of the measured observable.
This process is expected to be governed by the quantum theory of interaction between the measured particle, the measuring device, and the environment.
Before examining this process and integrating it with {\bf (RM)}, let us first address some general objections to the proposed model.

An immediate critique of the model arises from its reliance on the quantum dynamics, which is inherently linear and therefore incapable of breaking superpositions. Additionally, this approach appears to contravene established findings regarding the incompatibility of linear dynamics with the Born rule.
Let us first address this latter objection. It has been demonstrated in the paper that the Born rule for transition rates can be derived solely from the Schr{\"o}dinger equation with a Hamiltonian that satisfies {\bf (RM)}. This conclusion does not contradict \cite{nonLin} or other related publications, as those works incorporate additional assumptions, most notably stability (to be discussed separately), which inherently require some form of non-linearity. Consequently, the categorical assertion that linear evolution cannot yield the Born rule is inaccurate.

The notion that a linear transformation cannot break superpositions warrants further discussion. It is important to recognize that our measuring devices possess finite positional resolution, meaning they cannot differentiate state functions with sufficiently small support. Consider the ``squeezing" operator $A_{\lambda}$, defined on single-variable functions by $A_{\lambda}\varphi(x)={\sqrt \lambda}\varphi(\lambda x)$ for $\lambda>0$. This operator is linear and, in fact, unitary in $L_2(\R)$.
Now, suppose we have a superposition $\varphi = \alpha g_a + \beta g_b$, where $g_a$ and $g_b$ have small support and approximate the position eigenstates for points $a$ and $b$, respectively. By applying $A_{\lambda}$ to $\varphi$ and selecting a sufficiently large value of $\lambda$, we can reduce the interval containing the support of the resulting function $\psi$ to an arbitrarily small size. In such a scenario, a detector with finite resolution would be unable to distinguish $\psi$ from an eigenstate of the position. 
The feasibility of linear collapse models in scenarios where eigenstates are replaced by specific equivalence classes of states should be then re-examined.

As a side note, while not disputing the conclusion within accepted assumptions, it is worth noting that the derivation in \cite{nonLin} is limited to a two-dimensional state space.
The authors argue that any collapse model must be capable of describing the collapse of a two-state superposition, which is true. However, the fact that the initial state of a system is identified as a two-state does not necessarily imply that the evolution of the state under collapse will be restricted to a two-dimensional state space. Specifically, the two-space does not need to be an invariant subspace of the evolution operator.
For instance, when we measure the spin of a particle in the Stern-Gerlach experiment with a blocker, we utilize the entanglement between spin and position states and measure the particle's position along the field gradient. In this scenario, the appropriate collapse model may need to be formulated in the infinite-dimensional state space that accounts for both the spin and position of the particle.

Before examining the concept of stability in the model, let us first address what characterizes a macroscopic object according to conjecture {\bf (RM)}.
In this framework, a microscopic particle undergoing position measurement follows the trajectory dictated by the Schr{\"o}dinger equation with a stochastic Hamiltonian. The specific characteristics of the measuring apparatus are encapsulated by the distribution of elements in the Gaussian unitary ensemble random matrix representing the Hamiltonian, which evolve with time. 
Consequently, the particle's state undergoes a random walk in the space of states, with the step distribution being both homogeneous and isotropic.

    Assume the initial state of the particle lies in $M^{\sigma}_{3}$. By selectively sampling the steps of the random walk that take place on $M^{\sigma}_{3}$, we effectively simulate a random walk on $M^{\sigma}_{3}$ that approximates Brownian motion, described by the diffusion equation. As we scale up the size of the particle, the diffusion coefficient ${\mathbb D}$ associated with this equation diminishes, gradually approaching zero. We posit that the point at which ${\mathbb D}$ becomes practically negligible denotes a critical juncture, delineating the boundary between the macroscopic and microscopic realms in the framework. Moreover, when the particle's size is sufficiently large, its surroundings, including nearby particles and electromagnetic radiation, inherently harbor information about its position. The continuous measurement of position makes conjecture {\bf (RM)} applicable even without the need for a dedicated measurement device.
    When the particle is large enough that its diffusion coefficient becomes negligible for environmental measurements, perhaps limited to interactions with the cosmic microwave background, it may exhibit classical behavior naturally, rather than only under specific conditions such as in a bubble chamber.

 As an analogy, consider a free Brownian particle in $\R^3$ with no external potential. If the diffusion coefficient ${\mathbb D}$ is nonzero, the particle undergoes stochastic motion, and at any time $t$, the probability of finding it near a given point is described by a normal distribution.  When ${\mathbb D}$ is small, so that random environmental forces largely cancel, the particle remains effectively at rest in $\R^3$. If an external potential is introduced and damping is negligible, the system exhibits the classical behavior of a particle in a potential.

A similar situation arises with the state of a microscopic particle evolving under the rule {\bf (RM)}. When the diffusion coefficient ${\mathbb D}$ of the induced Brownian motion on $M^{\sigma}_3$ is nonzero, the state's evolution is stochastic, and the probability of reaching a given state in $M^{\sigma}_3$ follows the Born rule. However, when ${\mathbb D}$ approaches $0$, diffusion into the full state space ceases, and a state initially located on $M^{\sigma}_3$ remains there at rest - an example of Newtonian behavior for a body at rest. The condition ${\mathbb D} \approx 0$ is therefore essential; otherwise, the Hamiltonian in {\bf (RM)} would induce stochastic evolution that drives the state away from $M^{\sigma}_3$. We conclude that {\bf (RM)} is capable of explaining both the Born rule for position measurements of microscopic particles and the Newtonian behavior of macroscopic bodies at rest.

To derive the dynamics of macroscopic bodies, it is useful to begin with the decomposition of the state's velocity, given by $\frac{d \varphi}{dt}=-\frac{i}{\hbar}{\widehat h}\varphi$, where ${\widehat h}$ denotes the conventional Hamiltonian of a single particle in a potential ${\widehat V}$.  As demonstrated in \cite{KryukovNew}, the velocity of the state at an arbitrary point $\varphi$ in $M^{\sigma}_{3,3}$ in $\mathbb{CP}^{L_2}$ decomposes into three orthogonal components. The first two components replicate the classical velocity and acceleration of the particle, remaining tangent to the classical phase space manifold $M^{\sigma}_{3,3}$. The third component, orthogonal to the manifold, signifies the spreading velocity of the particle's state function. The squared norm of the state velocity in the Fubini-Study metric is thus the sum of the squares of these components, expressed by the following equation:
\begin{equation}
\label{decomposition}
\left\|\frac{d\varphi}{dt}\right\|^{2}_{FS}=\frac{{\bf v}^{2}}{4\sigma^{2}}+\frac{m^{2}{\bf w}^{2}{\sigma}^{2}}{\hbar^{2}}+\frac{\hbar^{2}}{32\sigma^{4}m^{2}},
\end{equation}
where ${\bf v}$ represents velocity, and ${\bf w}=-\frac{\nabla V}{m}$ denotes the acceleration of the particle.

Suppose now that an external potential ${\widehat V}$ is added to the Hamiltonian in {\bf (RM)}, and that ${\mathbb D}\approx 0$ for the particle. In this case, Newtonian dynamics emerges only under an additional assumption. The term ${\widehat V}$ pushes the state along $M^{\sigma}_{3,3}$ with acceleration $-\nabla V/m$, which is what we want. However, the corresponding path on $\R^3 \times \R^3$ is not Newtonian, as only the acceleration is accurately reproduced.  To make it Newtonian, we need to assume that the velocity and acceleration of the point in $\R^3 \times \R^3$ representing the  state moving in $M^{\sigma}_{3,3}$ correspond to the derivatives of position and velocity, respectively. That is, the manifold $M^{\sigma}_{3,3}$ may be identified with the classical phase space of the particle. In this case, the state's path $g_{a(t)}e^{ip(t)x/\hbar}$ describes Newtonian motion of a particle with position $a(t)$ and momentum $p(t)$. At any given time $t$ along the path, the squared norm of the state's velocity in the Fubini-Study metric is the sum of the first two terms in \eqref{decomposition}.

So, outside the case of a body at rest, one way to address the transition to classicality is to simply accept that the position, velocity, and acceleration of the  point on $\R^3 \times \R^3$ representing the motion of the state are related in the usual way. Of course, this means that, except in the case of rest, the evolution of the particle whose position is measured (including macroscopic particles ``measured" by the environment) is not described solely by the Hamiltonian in {\bf (RM)} plus potential. Additional terms are needed, and their existence must be demonstrated.

Suppose that we add the standard Schr{\"o}dinger Hamiltonian for the particle to the Hamiltonian in {\bf (RM)}, and assume that ${\mathbb D}\approx 0$. This removes the stochastic component of the evolution but reintroduces the issue of spreading.
Mathematically, the required additional term is a drift that counteracts the spreading and drives the state back toward $M^{\sigma}_{3,3}$, thereby ensuring that the last term in (\ref{decomposition}) vanishes. 
Assuming {\bf (RM)} and applying the macroscopicity condition ${\mathbb D} \approx 0$, we conclude that the drift process must be nearly deterministic, just like the spreading.
In Section IV, we will discuss the drift in more detail, including its potential physical origin and possible dynamical scenarios for both microscopic and macroscopic particles.

Now, let us talk about the stability assumption, a common aspect of discussions about the measurement problem. This assumption posits that following a measurement, a macroscopic measuring device should yield a definite outcome without spontaneously altering its readings. 
To investigate, let us consider a system of $n$ particles and the tensor product manifold $\otimes_{n} M^{\sigma}_{3}$ consisting of elements of the form $g_{1}\otimes...\otimes g_{n}$, where each $g_k \in M^{\sigma}_{3}$ represents the state of the $k$th particle. Similarly, we introduce the manifold $\otimes_{n} M^{\sigma}_{3, 3}$, which consists of tensor products of states of the particles in $M^{\sigma}_{3,3}$.
The transition from Schr{\"o}dinger to Newtonian dynamics, as explored for a single particle, naturally extends to systems comprising multiple interacting particles. In particular, a system of two particles whose state is constrained to the manifold $M^{\sigma}_{3,3}\otimes M^{\sigma}_{3,3}$ adheres to Newtonian dynamics.

Additionally, it is worth noting that the Euclidean metric on $M^{\sigma}_{3}$ extends to the Euclidean metric on the configuration space $\otimes_{n} M^{\sigma}_{3}=\R^{3n}$ of the particle system. This metric arises from the metric on the tensor product of Hilbert spaces of the particles' states. Furthermore, when an additional particle, described by a state $\varphi$ is considered alongside the $n$-particle system in $\otimes_{n} M^{\sigma}_{3}$, the product state $\varphi \otimes g_1 \otimes ... \otimes g_n$ of the total system is close to a state $g_a \otimes g_1 \otimes ... \otimes g_n$ in $\otimes_{n+1} M^{\sigma}_{3}$ under this metric precisely when $\varphi$ is close to $g_a$ in the Fubini-Study metric on the state space of a single particle. This observation will allow us to focus on the state of the particle rather than the state of the entire system when discussing the double-slit experiment later in the paper.
The same can be said about the classical phase space submanifold of an $n$-particle system.

Consider a system comprising a small classical particle interacting with a macroscopic measuring device. 
In Newtonian mechanics, the influence of the small particle on the measuring device can be disregarded, allowing the dynamics of the device to be treated independently. 
This independent of the particle interaction of the device with the environment result in its position being encoded in the environment, that is, measured. 
Consequently, conjecture {\bf (RM)} applies to the device alone. The Brownian motion on $\R^3=M^{\sigma}_3$, arising from the random walk in {\bf (RM)}, determines the probability distribution of the corresponding Hamiltonian entries, thereby uniquely defining the entire ensemble.
Given the device's macroscopic nature, its size surpasses the threshold for macroscopic behavior set by this interaction, rendering the induced Brownian motion trivial. Thus, its state in the rest system, $g_{D{\bf a}}$, belongs to the submanifold $M^{\sigma}_{3}$ (or $\otimes_n M^{\sigma}_{3}$, if the device's components are considered) of the device's state space and remains unchanged.
With the prior identification of $M^{\sigma}_{3,3}$ as the classical phase space, or alternatively, by incorporating drift as discussed in Section IV, introducing an external potential to the system results in its behavior being governed by Newtonian dynamics.

By embracing conjecture {\bf (RM)}, we have obtained knowledge of the state of the measuring device during measurement. Consequently, the state of the particle-device system during measurement in the model must be a product state. As long as the small particle itself is macroscopic, it also follows Newtonian dynamics and evolves in the potential created by the device. In this case, the particle-device system exists in a product state $g_{P{\bf a}, \sigma}\otimes g_{D{\bf a}, \sigma}$, where $g_{P{\bf a}, \sigma}$ in $M^{\sigma}_{3,3}$ denotes the state of the particle. 
If the particle's size decreases to fall below the macroscopic threshold, its state ``detaches" from the manifold $M^{\sigma}_{3,3}$ and, during measurement, undergoes a random walk on the space of states.
The probability of reaching various position eigenstates during this walk conforms to the Born rule. In both scenarios, the system's state in the model remains a product, and there's never a moment where the device and the particle form a ``cat-state."

Traditionally, the examination of state measurement begins with an entangled state involving a microscopic entity and the measuring apparatus. The challenge lies in demonstrating how this entangled state produces definite measurement outcomes during the measurement process. However, under the conjecture {\bf (RM)}, the measurement process unfolds differently from the outset. Here, entanglement exists solely between microscopic entities. Throughout the measurement, the macroscopic measuring device and the measured particle exist in a product state. At all times, the device satisfies the ${\mathbb D}\rightarrow 0$ limit and follows Newtonian dynamics. Consequently, the stability of measurement outcomes recorded by the macroscopic device in the model is inherently guaranteed by Newtonian dynamics and the minimal impact of the environment on the device that follows from it.

In this framework, what characterizes the collapse of the state of the measured particle under a position measurement?
The interaction between a microscopic particle and the device induces a random walk of the particle's state, governed by the Schr{\"o}dinger equation with the Hamiltonian in {\bf (RM)}. Throughout the measurement, the particle-device system remains in a product state, ensuring that the particle's state by itself is defined, at least in principle.
The collapse 
occurs when the state assumes the form of a state $g_{{\bf a}, \sigma}$ or one of the states in its equivalence class, defining a point in $M^{\sigma}_{3}$. As is known, subsequent measurements in this case will yield the same result if conducted immediately afterward, indicating that no further collapse is occurring. Additionally, the presence of the measuring device at the time the state takes this form is irrelevant. 
In particular, recording the measurement result merely confirms that the state has attained the required form and does not contribute to the collapse.

It has been established that the conditional probability for the state of a measured particle to arrive at a specific point on the manifold $M^{\sigma}_{3}$, given that the state has reached the manifold, follows the Born rule. However, since the distribution of steps in the random walk of the measured particle's state is isotropic, we still need to explain why the state is capable of reaching the manifold $M^{\sigma}_{3}$.
As demonstrated in Section III, this can be achieved using equivalence classes of eigenstates indistinguishable to the detector. Alternatively, introducing a drift term into the state's random walk, as also shown in Section III and validated by computer simulations, provides another approach.
However, the challenge, not fully addressed in the paper, lies in identifying the physical origin of this drift (see also Section IV).

The collapse process in the model with a drift is effectively represented by two distinct processes, both assumed to arise from the interaction between the measured particle and the device: the evolution driven by a random matrix Hamiltonian and the narrowing of the state function.
The former ensures that the Born rule is satisfied for position measurements, and potentially for other types of measurements as well.
The role of the second process is to drive the state of the measured system towards the manifold of equivalence classes of position eigenstates without interfering with its orthogonal motion {\it along} the manifold, which is responsible for the Born rule.
Devices designed to measure different observables are distinguished by how the second of these processes is correlated with the measured observable. While the position is measured in any device, its association with the observable in question varies across different devices. For instance, when measuring the momentum of a charged particle, one can employ a magnetic spectrometer to bend the particle's path and subsequently measure its position on a screen to deduce its momentum during motion.

Here, we apply the conjecture  {\bf (RM)} to analyze a ``which-way" type of measurement in the double-slit experiment. We re-derive the Born rule, which, in a specific case, emerges from a simple ``gambler's ruin" random walk. We then provide physical and mathematical insights into the evolution of the state driven by the Hamiltonian in {\bf (RM)}, both with and without drift, and present computer simulations of the model.
The path of the state between the source, the screen with the slits, the detector, and the backstop screen is traced.
It is demonstrated that the space of states and the Fubini-Study metric on it provide a suitable framework for the experiment, shedding new light on its mysteries. The general results presented in this section and in \cite{KryukovNew} are made more tangible and useful 
for understanding the process of collapse in this fundamental case.

\section{The double-slit experiment with a measurement}

Consider the double-slit experiment with a microscopic particle of mass $m$ whose motion is adequately described by the Schr{\"o}dinger equation. Let us choose the $z$-axis on the screen with the slits, orthogonal to the slits. Suppose the $z$-coordinates of the lower and upper slits are $a$ and $b$, respectively. Let the horizontal axis run along the particle's path from left to right, as shown in Figure 1.

\begin{figure}[ht]
\centering
\includegraphics[width=8cm]{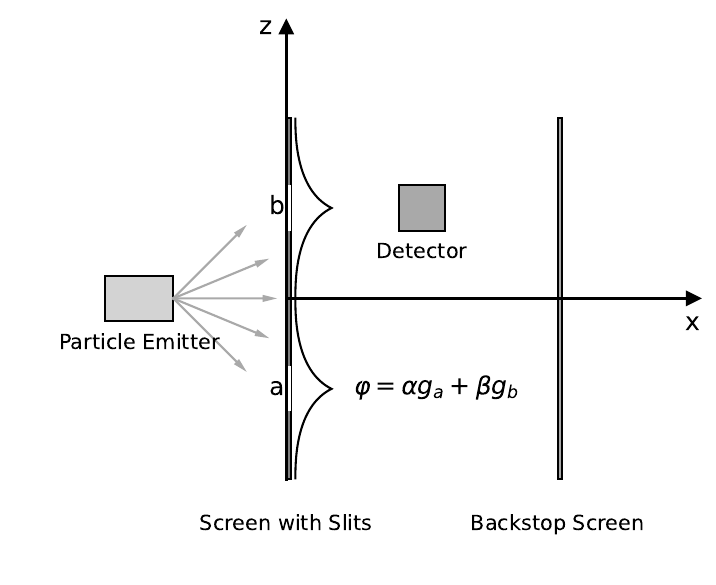}
\caption{Double-slit experiment with a measurement.}
\label{fig:1}
\end{figure}

At a point immediately to the right of the slits, the particle is in a superposition of states $g_a$ and $g_b$, representing the particle passing through one of the slits with the other slit closed. The state of the particle at that point can be identified with a function $\varphi=\alpha  g_a+\beta  g_b$, where $\alpha$ and $\beta$ are complex constants.
For this paper, the functions $g_a$ and $g_b$ immediately to the right of the slits can be approximated by Gaussian functions of $z$ of a certain ``width" $\delta$, peaked at $a$ and $b$ respectively. Interaction of the particle with the screen is described by the usual Schr{\"o}dinger equation. Thus, the Schr{\"o}dinger evolution takes the initial state of the particle at the source to the two-dimensional space of states $\C^2$ of linear combinations of $g_a$ and $g_b$, or, more precisely, to the projective space $\mathbb{CP}^1=S^2$ formed by the unit states in $\C^2$ modulo the phase factor.

Let us now insert a particle detector by one of the slits on the right. By measuring the particle's position, the detector provides information about the slit near which the particle is located at the time of measurement.  This is an example of what is called the ``which way" measurement. To make the measurement successful, we need to assume that $|a-b| \gg \delta$, so that the states $g_a$ and $g_b$ are nearly orthogonal. In fact, if the ``overlap" of $g_a$ and $g_b$ is significant, no detector will be able to identify the slit by which the particle is located. In particular, the detector should be placed sufficiently close to the screen, before $g_a$ and $g_b$ spread and start interfering. With this in place, the measurement causes the collapse of the wave function and results in a transition from wave to particle properties of the system. The common view is that the measurement tells us which slit the particle went through.

For simplicity and to be specific, let us assume the detector is a small scintillation screen positioned near the slit at $z=b$. The detector's role is to confirm or deny the particle's location by the slit at the time of observation.  Let the state function of the particle detected at a point of the scintillation screen be denoted by $\eta$. Realistically, $\eta$ cannot be the Dirac's delta function; its support must be at least the size of the scintillator material's atom on the screen. We divide the screen into cells of the corresponding small size $d_\eta$ and identify the state of the particle detected in the $k$-th cell by the normalized characteristic function $\eta_k$ of the cell. An ideal detector would detects the particle in a state $\eta_k$ with probability $1$. 
The probabilities $P_b=\sum_{k}|(g_b, \eta_k)|^2$ and $P_a=\sum_{k}|(g_a, \eta_k)|^2$, where $( \ , \  )$  denotes the inner product between states, quantify the effectiveness of an ideal detector in the experiment. 
These probabilities depend on the functions $g_a$ and $g_b$ as well as the position, size, and the ``granularity" parameter $d_\eta$ of the detector. 
Note that $P_b$ must be sufficiently high, and $P_a$ must be sufficiently small to identify the slit by which the particle was found. We then say that the particle in state $\varphi$ is near the slit $z=b$ if $\sum_{k}|(\varphi, \eta_k)|^2 \ge P_b-\epsilon$ for some $\epsilon>0$, sufficiently small for the state to identify the slit. 
 This condition is met by a range of states $\varphi$ that are all identified with $g_b$ in the experiment. The resulting equivalence class of states will be called the {\it physical eigenstate} of the position operator on the $z$-axis. In this case, we will also say that the state $\varphi$ is {\it measurable without dispacement} by the detector. 

Note that for functions $\varphi$ that do not vary significantly across the cells, the term $\sum_k |(\varphi, \eta_k)|^2$ in the definition of a physical eigenstate is approximately equal to the squared norm of the component $\varphi_D$ of $\varphi$, obtained by setting it to zero outside the interval $D$ occupied by the detector.
 If $\eta_D$ is the characteristic function of $D$, then $\varphi_D=\varphi \cdot \eta_D$. The state $\varphi$ is in the equivalence class of $g_b$ if the ``tails" of $\varphi$ outside $D$ are sufficiently small, i.e., the norm of $\varphi-\varphi_D$ is small.  For an arbitrary value of $c$ in $z$, the equivalence class $\{g_c\}$ of the state $g_c$ is defined in the same way as for the class $\{g_b\}$, by translating the interval $D$.  
Note that a state in the equivalence class $\{g_a\}$ of $g_a$ is approximately orthogonal to a state in the equivalence class $\{g_b\}$ of $g_b$. In what follows, we will assume that such orthogonality of states is fulfilled.

 Let us define the Fubini-Study distance between a state $\varphi$ and the equivalence class 
$\{g_b\}$ by
\begin{equation}
\label{dist}
\rho(\varphi; \{g_b\})=\inf_{\psi \in \{g_b\} } \rho(\varphi; \psi),
\end{equation} 
where $\rho(\varphi; \psi)$ is the Fubini-Study distance between states. In particular, for the distance between $\varphi=\alpha g_a+\beta g_b$ and $\{g_b\}$ under the accepted conditions, we have   $\cos \rho(\varphi; \{g_b\})=|\beta|$. 
For the state $\varphi$ to reach the physical eigenstate $\{g_b\}$, it is necessary and sufficient that $\rho(\varphi; \{g_b\})=0$. 
Note that the equivalence class $\{g_b\}$ of the eigenstate $g_b$ is rather ``large". In particular, it contains functions with support in the interval $D$ occupied by the detector, provided their total variation is not too large. It follows that $\{g_b\}$ contains many orthogonal states, i.e., states at the Fubini-Study distance equal to the maximal possible value of $\pi/2$ from each other. 
 
 To clarify the role of the equivalence class during a measurement, let us consider a few examples. 
 We assume
a slit separation of $10^{-5}m$, a slit-width of $10^{-9}m$, and that the width parameter $\delta$ of the states $g_{a}$ and $g_b$ is comparable to the slit-width. These values are typical for a successful experiment of this sort. The length of the detecting scintillation screen by the slit is taken to be about half the slit separation.
Suppose the initial state $g_b$, denoted as $g_{b, \delta}$ here, moves to the point represented by the Gaussian state $g_{b, 100\delta}$ with a width of $100\delta$. Taking the inner product of the states yields $|(g_{b, \delta},g_{b, 100\delta})|=\cos \rho$, where $\rho=\rho(g_{b, \delta},g_{b, 100\delta})$ denotes the Fubini-Study distance between the states. We then have $\rho \approx 1.43$ radians or about $82^{\circ}$. Because $100\delta=10^{-7}m$, the width of the state $g_{b, 100\delta}$ is less than the size of a scintillation screen. In particular, the condition $\sum_{k}|(g_{b, 100\delta}, \eta_k)|^2 \ge P_b-\epsilon$ is satisfied for a small $\epsilon$. It follows that the state $g_{b, 100\delta}$ is still within the equivalence class of $g_b$, and thus, it represents the same physical eigenstate. On the other hand, we also have $|(g_{a, \delta}, g_{b, 100\delta})|<\exp(-10^4)$, which is an extremely small number. So, by any measure the states $g_{a, \delta}$ and  $g_{b, 100\delta}$ can be considered orthogonal, as needed for the experiment.   

For the second example, consider that the state $g_b=g_{b, \delta}$ is displaced by a distance of $10\delta=10^{-8}$ along the $z$-axis. We then have $|(g_{b, \delta},g_{b-10^{-8}, \delta})|<\exp(-12)$, corresponding to a Fubini-Study distance of about $89.999^{\circ}$. So, the states are nearly orthogonal. However, because $10^{-8}$ is much smaller than the size of the detector, the condition $\sum_{k}|(g_{b-10^{-8},\delta}, \eta_k)|^2 \ge P_b-\epsilon$ is satisfied for a small $\epsilon$. It follows that the states $g_{b, \delta}$ and $g_{b-10^{-8}}$ belong to the same equivalence class. At the same time, the states $g_{a, \delta}$  and $g_{b-10^{-8}}$ remain orthogonal to a very high degree of accuracy, as required for successful measurement.

Suppose now that the initial state is a superposition $\varphi=\alpha g_a+\beta g_b$ with moduli $|\alpha|$ and $|\beta|$ that are away from zero, for example, $\varphi=\frac{1}{\sqrt 2}g_a+\frac{1}{\sqrt 2} g_b$. Unlike the states $g_a$ and $g_b$, the state $\varphi$ cannot be ``measured without displacement" by the detector capable of resolving the slits. In other words, such a state does not satisfy the condition $\sum_{k}|(\varphi, \eta_k)|^2 \ge P_b-\epsilon$ with a small $\epsilon$ or a similar condition for the detector located at $z=a$. In other words, the superposition $\varphi$ is far from the physical eigenstates of the measured particle. The measurement happens only if and when the initial state $\varphi$ is moved to the equivalence class of either $g_a$ or $g_b$. The Fubini-Study distance from the state $\varphi=\frac{1}{\sqrt 2}g_a+\frac{1}{\sqrt 2} g_b$ to $\{g_b\}$ is 
\begin{equation}
d(\varphi; \{g_b\})=
\frac{\pi}{4}rad.
\end{equation} 
So, the initial state $\varphi$ traveling the distance  of $\pi/4$ along the shortest geodesics towards the physical eigenstate $\{g_b\}$ will reach the physical eigenstate and become directly measurable by the detector. 
At the same time, 
the state $\varphi=\alpha g_{a+10^{-8}, \delta}+\beta  g_{b-10^{-8}, \delta}$ based on the earlier example travels almost twice the distance from the initial state $\alpha g_{a, \delta}+ \beta g_{b, \delta}$ but is still 
at the same distance from the physical eigenstate $\{g_b\}$. 
The reason for the difference between the first two and the last example is due to the fact that the detector stretches along interval $D$ in the $z$-axis.
This makes displacements in $D$ or relatively small changes in the width parameter of $g_b$ possible without affecting the distance of the resulting state to the equivalence class $\{g_b\}$.

Let us return to the double-slit experiment with both slits open and the detector near the slit $z=b$. 
According to {\bf (RM)}, the observed state $\varphi$ will be acted upon by the Hamiltonian represented by a random matrix and will perform a random walk on the space of states. As a result of this walk, the state may be able to reach one of the physical eigenstates of the measured observable. 
Our main goal is to find the probability of transition of the initial state to physical eigenstates $\{g_a\}$ and $\{g_b\}$ for this experiment. Additionally, 
because the distribution of steps of the random walk of the state is isotropic and the space of states $\mathbb{CP}^{L_2}$ is infinite-dimensional, we need to ensure that the probability of reaching an eigenstate is non-vanishing to begin with.

To achieve these goals, let us utilize the expected value $\mu_z$ and the standard deviation $\delta_z$ of the $z$-coordinate to identify a submanifold of $\mathbb{CP}^{L_2}$ helpful for describing the measurement and to establish a coordinate system on it. We have:
\begin{equation}
\label{mu}
\mu_z=\int z |\varphi(z)|^2 dz,
\end{equation}
and 
\begin{equation}
\label{sigma}
\delta^2_z=\int (z-\mu_z)^2 |\varphi(z)|^2dz.
\end{equation}
Given an initial state $\varphi$ with a finite expected value $\mu_z$ and standard deviation $\delta_z$, consider the two-dimensional manifold $M_\varphi$ parametrically defined by
\begin{equation}
\label{surf}
\varphi_{\tau,\lambda}(z)={\sqrt \lambda}\varphi(\lambda (z-\mu_{z}-\tau)+\mu_z).
\end{equation}
The numeric parameters $\tau$ and $\lambda$ serve as coordinates on the manifold. Along the path $\varphi_\tau=\left.\varphi_{\tau, \lambda}\right |_{\lambda=\lambda_0}$ with a fixed value of $\lambda$, the expected value changes from $\mu_z$ to $\mu_z+\tau$, while the standard deviation remains constant. Similarly, along the path $\varphi_\lambda=\left.\varphi_{\tau, \lambda}\right |_{\tau=\tau_0}$ with fixed $\tau$, the standard deviation changes from $\delta_z$ to $\delta_z/\lambda$, while the expected value stays the same.

The motion along $\varphi_\lambda$ ``squeezes" or ``stretches" the state function without altering its shape or translation. This motion can relocate the state from its initial position in the state space $\mathbb{CP}^{L_2}$ to the $z$-axis represented by the family of equivalence classes $\{g_c\}$, where $c=\mu_z$ lies on the $z$-axis. Similarly, motion along $\varphi_\tau$ translates the state along the $z$-axis. This motion can bring the ``squeezed" state to the detector. The role of the equivalence class is crucial in this process: squeezing a state may not move it closer to a $g_c$-state by itself, but it will bring it closer to an equivalence class $\{g_c\}$.

Unlike the Fubini-Study distance between states, the expected value $\mu_z$ and standard deviation $\delta_z$ have the advantage in being familiar spatial quantities. Moreover, the condition that the initial state $\varphi$ has reached the detector or, equivalently, that it became a physical eigenstate of $z$ can be expressed in terms of the corresponding change in the variables $\mu_z$ and $\delta_z$ of $\varphi$. Specifically, for this to happen, it is sufficient that the interval $(\mu_z-r\delta_z, \mu_z+r\delta_z)$ for a proper value of the parameter $r$ for the final state $\varphi_f$ is contained in the interval $D$ occupied by the detector. First, for the given values of $\delta_z$ and $\mu_z$ of the initial state $\varphi$, the parameter $r>0$ is selected to ensure that the tails of $\varphi$ outside the interval  $D_r=(\mu_z-r\delta_z, \mu_z+r\delta_z)$ are small enough to satisfy the condition  $\Sigma_{k}|(\varphi, \eta_k)|^2 \ge P-\epsilon$ on the interval. Then, the coordinates $\tau$ and $\lambda$ (i.e., the corresponding values of $\mu_z$ and $\delta_z$) are selected to make sure that the interval  $(\mu_z-r\delta_z, \mu_z+r\delta_z)$ is in $D$. The range of possible values of $\mu_z$ and $\delta_z$ that satisfy this condition determines the end-states $\varphi_f$ in $M_\varphi$ that are elements in the corresponding physical eigenstate of $z$.

While collapse can be modeled using the variables $\mu_z$ and $\delta_z$ for more general states, we will, for now, focus on superpositions $\varphi = \alpha g_a + \beta g_b$, which are relevant to the double-slit experiment. As discussed, these state functions can be viewed as elements of the space $\C^2$, though operations such as ``squeezing" and translating a function act in an infinite-dimensional function space.
The manifold $M_\varphi$ forms an overcomplete frame (basis) in the space $L_2(\R)$, analogous to the manifold $M^{\sigma}_{3} = \R^3$, and it will similarly be used to establish a connection between the normal probability distribution and the Born rule.
Under {\bf (RM)}, the evolution of a quantum system's state in infinite-dimensional state space provides the most general approach to collapse. We will demonstrate that the obtained results also apply to spin-state collapse.

Let us prove that the steps of the random walk of the state $\varphi=\alpha g_a+\beta g_b$ along the paths $\varphi_{\tau}$ and $\varphi_{\lambda}$ on $M_\varphi$ are independent random variables. As we know, the probability distribution of the random vector representing a step in {\bf (RM)} is a normal isotropic distribution. The orthogonal components of such a vector are independent random variables. Therefore, we need to check that the steps along these paths take place in the projective space of states and that they are orthogonal in the Fubini-Study metric. Let us first check that this is true for the steps originating at the initial state $\varphi=\alpha g_a +\beta g_b$. First of all, because the norm of the state along the paths $\varphi_{\tau}$ and $\varphi_\lambda$ is preserved, the paths take values on the unit sphere $S^{L_2}$ in the space of states. In particular, the vectors $\frac{d \varphi_\lambda}{d\lambda}$ and $\frac{d \varphi_\tau}{d\tau}$ are tangent to the sphere. Also, $\left. \frac{d\varphi_\tau}{d\tau} \right|_{\tau=0}=-\frac{d \varphi}{dz}$ and $\left. \frac{d\varphi_\lambda}{d\lambda} \right|_{\lambda=1}=\frac{1}{2}\varphi+\frac{d \varphi}{dz}(z-\mu_z)$, and for the state $\varphi=\alpha g_a+\beta g_b$ we have
\begin{equation}
\mathrm{Re}\left(i\varphi, -\frac{d \varphi}{dz}\right)=0
\end{equation}
\linebreak
and 
\begin{equation}
\mathrm{Re}\left(i\varphi, \frac{1}{2}\varphi+\frac{d \varphi}{dz}(z-\mu_z)\right)=0
\end{equation}
by the properties of states $g_a$ and $g_b$. It follows that the vectors  $\frac{d \varphi_\lambda}{d\lambda}$ and $\frac{d \varphi_\tau}{d\tau}$ are orthogonal to the fibre of the fibration $S^{L_2} \longrightarrow \mathbb{CP}^{L_2}$. In particular, they can be thought of as vectors tangent to the projective space of states $\mathbb{CP}^{L_2}$. 
Now,
\begin{equation}
\label{prom}
\mathrm{Re}\left(\left. \frac{d\varphi_\lambda}{d\lambda} \right|_{\lambda=1}, \left. \frac{d\varphi_\tau}{d\tau} \right|_{\tau=0}\right)=-\mathrm{Re}\left(\frac{d\varphi}{dz}, \frac{1}{2}\varphi+\frac{d \varphi}{dz}(z-\mu_z)\right).
\end{equation}
Using the orthogonality of $\varphi$ and $\frac{d \varphi}{dz}$, the approximate orthogonality of $g_a$, $g_b$, and their derivatives, the equality $(g_a, \frac{d^2 g_a}{dz^2}) = (g_b, \frac{d^2 g_b}{dz^2})$, along with the expression for $\varphi$ and the definition of $\mu_z$, the resulting expression (\ref{prom}) can be rewritten and evaluated as follows:
\begin{widetext}
\begin{eqnarray}
\mathrm{Re}\left(\varphi \cdot (z-\mu_z), \frac{d^2 \varphi}{dz^2}\right) = \left(|\alpha|^2 (a-\mu_z)+|\beta|^2(b-\mu_z)\right) \left(g_a, \frac{d^2 g_a}{dz^2}\right)=0.
\end{eqnarray}
\end{widetext}
This proves the orthogonality of steps from the initial state $\varphi$ along the paths $\varphi_{\tau}$ and $\varphi_{\lambda}$. The application of the chain rule demonstrates that the preceding calculations remain valid for steps from any point on $M_\varphi$.

The established orthogonality confirms that steps of the random walk from any state $\psi$ in $M_\varphi$ along the direction tangent to paths $\varphi_\tau$ and $\varphi_\lambda$ through $\psi$ are independent random variables. Furthermore, it is possible to re-parametrize the paths $\varphi_\lambda$ to make the Fubini-Study metric on $M_\varphi$ in the new coordinates explicitly Euclidean. Specifically, by setting $s=\ln \lambda$, we obtain the new parametrization of $\varphi_\lambda$ in the form $\varphi_s(z)=e^{\frac{s}{2}}\varphi(e^s (z-\mu_z-\tau_0)+\mu_z)$. We can see that the norm of the tangent vector $\frac{d \varphi_s}{ds}$ is preserved along the path. The same is true for $\frac{d \varphi_\tau}{d\tau}$, which, together with the orthogonality of these vectors, signifies that the induced metric is Euclidean. The coordinates $\tau$ and $s$ are then Cartesian coordinates on $M_\varphi=\R^2$.

An arbitrary state on $M_\varphi$ has the form  $\psi=\alpha \widetilde{g}_{c}+ \beta \widetilde{g}_{d}$, where $\widetilde{g}_{c}$ and $\widetilde{g}_{d}$ are Gaussian functions with equal width, and $(\widetilde{g}_{c}, \widetilde{g}_{d})=(g_a, g_b)$. 
The expected value of the $z$-coordinate for an arbitrary state $\psi$ in $M_\varphi$ is given by
\begin{equation}
\label{z}
\mu_z=\int z |\alpha \widetilde{g}_{c}+ \beta \widetilde{g}_{d}|^2 dz
=|\alpha|^2 c+|\beta|^2 d.
\end{equation}
The variance is given by
\begin{equation}
\label{sigg}
\delta^2_z=\int z^2  |\alpha  \widetilde{g}_{c}+\beta \widetilde{g}_{d}|^2 dz-\mu^2_z=|\alpha|^2|\beta|^2 (c-d)^2.
\end{equation}
Provided the coefficients $\alpha$ and $\beta$ do not vanish, equations (\ref{z}) and (\ref{sigg}) can be solved for $c$ and $d$. If one of the coefficients is $0$, the state is an eigenstate of $z$. In either case, we see that the pair $(c,d)$ for the states on $M_\varphi$  can be represented by the pair  $(\mu_z, \delta_z)$, identified in this context with coordinates $\tau$ and $s$. It follows that the Fubini-Study distance from a state in $M_\varphi$ to the eigenstates $g_a$ and $g_b$ can be expressed through the values of $\mu_z$ and  $\delta_z$ for the state. The Fubini-Study distance $d \rho$ between two neighboring points of  $M_\varphi$ can be expressed through the differentials $d \tau$ and $d s$ for the points as follows:
\begin{equation}
d \rho ^2=d \tau^2+d s^2.
\end{equation}

As discussed in Section II, the motion of states under the Schr{\"o}dinger equation with a Hamiltonian in {\bf (RM)} results in Brownian motion that satisfies the diffusion equation on the classical space submanifold in the space of states.
In the case of state functions depending on a single variable $z \in \R$, the Hilbert space of states is $L_2(\R)$, and the classical space submanifold, denoted here as $M^{\delta}_1$, is $\R$. 
The point source solutions to the diffusion equation are 
single-variable Gaussian functions $\rho_t$. The classical space manifold comprises Gaussian functions $g_c = \sqrt{\rho_t}$ for a fixed value of $t$, corresponding to the width parameter $\delta$ of $g_c$. This manifold inherits the Euclidean metric from the Fubini-Study metric on the projective space of states $\mathbb{CP}^{L_2}$, making it isometric to $\R$.

By altering the state space of the system, we obtain Brownian motion on $\R$, which satisfies the diffusion equation with suitable boundary conditions. Although the distribution of steps in the random walk that approximates the Brownian motion will change, for small steps it remains approximately normal. Thus, the random walk that satisfies the imposed boundary conditions can still be considered as having Gaussian steps. Alternatively, one can substitute functions $g_c$ with functions $r_c$ in the equivalence class, whose square yields the density $\rho_t$  for a fixed $t$. In either case, the Born rule will be valid on $\R$ or the appropriate interval thereof, and therefore on the entire space of states.

Let us use the manifold $M_\varphi$ to illustrate the process of collapse and the emergence of the Born rule from the random walk associated with {\bf (RM)}, both in cases where the boundary conditions are imposed and where they are not.
Suppose first that the random walk of the initial state $\varphi=\alpha g_a+\beta g_b$ generated by the Hamiltonian ${\widehat h}$ in {\bf (RM)} takes place on the manifold $M_\varphi$. That is, we select only those steps of the walk generated by ${\widehat h}$ that begin and end on $M_\varphi$. We will address the known isotropy of the distribution of steps later. Note that the states $g_a$ and $g_b$ are the points of $M_\varphi$ where $\mu_z=a$ or $\mu_z=b$ and $\delta_z=\delta$. 
The classical space manifold $M^{\delta}_{1}$, representing the $z$-axis in the Hilbert space of states $L_2(\R)$, will consist of equivalence classes of states in $M_{\varphi}$. Each class comprises functions $\psi$ with a fixed expected value $\mu_z$ and a standard deviation $\delta_z$ satisfying $\delta_z \leq \delta$. The equivalence classes represent physical eigenstates of $z$.

For our first example, no boundary conditions are imposed. From the properties of ${\widehat h}$ in {\bf (RM)} and from the isometry between $M_\varphi$ and $\R^2$, we infer that the random walk of the state on $M_\varphi$ is a random walk with Gaussian steps on $\R^2$. The coordinates $\tau$ and $s$ are orthogonal and the steps in $\tau$ and $s$ are independent, identically distributed normal random variables. 
It follows that the probability density function of the random vector of the final state $\varphi_f$ at the time of observation is a normal, circularly symmetric function of $\tau$ and $s$ on $\R^2$. Therefore, the probability of the particle being located near $a$ or $b$ is the product of the probability that the expected value $\mu_z$ is near $a$ or $b$ and the probability that the standard deviation $\delta_z$ is less than $\delta$.
However, for a given initial state, the probability of $\delta_z$ being less than $\delta$ is just a constant coefficient, which is the same for the convergence of the initial state to either $g_a$ or $g_b$. In other words, the probability we are considering is proportional to the probability of $\mu_z$ being near $a$ or $b$ on the $z$-axis. 

From $\varphi_\tau(z)=\varphi(z-\tau)$, we have $d\tau=d\mu_z=-dz$, so that we are dealing with a random walk on the $z$-axis.
This random walk approximates Brownian motion, which solves the diffusion equation on $\R$. 
 If function $\psi$ represents a point on the $z$-axis, then, in the given approximation, $|\psi|^2 = |\alpha|^2 {\widetilde g}_c^2 + |\beta|^2 {\widetilde g}_d^2$.
 The function $|\psi|^2$ serves as a solution to the diffusion equation at a fixed time, given the initial condition of two point sources located at two nearby points $c$ and $d$ on the $z$-axis. It follows that the Born rule, when applied to the states constituting $M^{\delta}_{1}$, yields the probability density function that solves the diffusion equation. Conversely, because the Fubini-Study distance between states in $M^{\delta}_{1}$ spans all values from $0$ to $\pi/2$, and because the probability of transition between any two states linked by the random walk specified in {\bf (RM)} depends solely on this distance, the validity of the Born rule for all states can be inferred from its validity on $M^{\delta}_{1}$ (see Section II). Also note that as the parameter $\delta_z$ approaches $0$, the initial condition of two point sources converges to the delta function. Consequently, we could equally well use Gaussian functions to represent equivalence classes of points on the $z$-axis. In particular, this derivation closely parallels the one leading to equation (\ref{distt}) in Section II.

For a more ``visual" derivation of the Born rule from the state walk on the manifold $M_\varphi$,
let us note that in the considered approximation, the states $g_a$ and $g_b$ are orthogonal, indicating that they occupy opposite points in the state space.
It follows that in this approximation, the expected value $\mu_z$ of the coordinate $z$ cannot exceed the value $b$ or be smaller than $a$. 
It also follows that there is a maximum possible value of the standard deviation $\delta_z$ of $z$. According to (\ref{sigg}), this value is equal to $|\alpha||\beta||a-b|$. These constraints are acceptable because there is a very small probability for the particle to be found beyond a small neighborhood of the interval $[a,b]$, which separates the slits in a properly set up experiment. In effect, we place the particle in a box. Our second example of deriving the Born rule in the model will use these constraints as boundary conditions for a random walk on the interval $[a, b]$. 
As previously discussed, under {\bf (RM)}, the random walk on the interval $[a,b]$ alone already ensures the Born rule on the state space $L_2[a,b]$. The random walk in the standard deviation variable (without drift) serves primarily as an illustration of the squeezing of the state during the collapse process in the model. 

Let us impose absorbing boundaries at $\tau = a$ and $\tau = b$. To ensure the product form of the joint probability and satisfy the constraint on $\delta_z$, we also impose a reflecting boundary condition at $\delta_z = |\alpha||\beta||a - b|$. 
The required probability is the product of the probability that $\mu_z$ is near $a$ or $b$, and the probability that $\delta_z$ is less than $\delta$. Once again, the probability that $\delta_z < \delta$ is a constant factor, identical for the convergence of the initial state to either $g_a$ or $g_b$.
It follows that the problem of finding the probability of transition of the initial state to $g_a$ or $g_b$ can be solved by studying 
the random walk with Gaussian steps on the interval $[a,b]$.
When the number of steps is large, the obtained walk with Gaussian steps can be approximated by the walk whose steps have a fixed length.
 The end-points of the interval $[a, b]$ are absorbing. 
 The probability of reaching the point  $\mu_z=b$ for the state  $\varphi=\alpha g_a+\beta g_b$ is then given by the usual gambler's ruin formula that  yields in this case
\begin{equation}
\label{born}
P_{b}=\frac{ \textrm{number of steps from} \ \mu_z \  \textrm{to}  \ a}{\textrm{number of steps from} \ a \ \textrm{to} \ b}=
\frac{\mu_z-a}{b-a}=|\beta|^2.
\end{equation} 
Here the definition (\ref{z}) together with normalization $|\alpha|^2+|\beta|^2=1$ were used. Similarly, the probability $P_{a}$ for the initial state $\varphi$ of reaching the state $g_a$ (equivalently, reaching $\mu_z=a$)
is given by 
$P_{a}=|\alpha|^2$. 
The Born rule is thus derived for the constrained state in $M_\varphi$, and, consequently, for arbitrary states in the state space.

Let us emphasize that the collapse of the state in the model is associated with the state approaching a physical eigenstate of position during the random walk. The state becomes well-localized, positioned at either point $a$ or $b$, indicating that a collapse has occurred. 
Although the presence of absorbing boundaries in the second example alters the walk by changing the space of states, the absorption process itself is not equivalent to collapse.
In particular, the state becomes well-localized as it approaches the endpoints {\it before} being absorbed.
The measuring device in the model influences the state's evolution by subjecting it to a random walk, as described in {\bf (RM)}. However, the device's role after the state reaches $g_a$ or $g_b$ is simply to register this outcome. Moreover, whether or not the outcome has been recorded, the collapse has already occurred. 
As explained in Section II, stability of a macroscopic measuring device in this framework is ensured by the transition to Newtonian dynamics in the ${\mathbb D}\rightarrow 0$ limit. Since the Born rule for the particle's state is also satisfied, collapse in the framework is equivalent to a dynamic approach to a physical eigenstate of position.

The random walk of the state in the model was constrained to remain on the manifold $M_\varphi$. Even with this constraint, reaching the eigenstates $g_a$ and $g_b$ depends on the standard deviation $\delta_z$ falling below the value $\delta$, associated with the size of the measuring component of a detector. 
The isotropic distribution of state steps, driven by the Hamiltonian in {\bf (RM)}, enables propagation into the space of states $\mathbb{CP}^{L_2}$, seemingly reducing further the likelihood of reaching an eigenstate. This issue can be addressed mathematically either by adding a drift term to the random walk in ${\bf (RM)}$ or by considering equivalence classes of states that are indistinguishable to the measuring device. 
In the latter case, the initial state evolving under {\bf (RM)} can still reach the equivalence classes $\{g_a\}$ and $\{g_b\}$ with probabilities consistent with observations.

To describe the propagation of the state into $\mathbb{CP}^{L_2}$, we must consider superpositions of more than two Gaussian states. 
It is known that the set of finite linear combinations of translations of a single Gaussian function is dense in $L_2(\R)$ (see, for example, \cite{arx}). This observation is important when working with superpositions of Gaussians and also provides meaningful context for the foliation construction in state space that follows. 
At the same time, the assumption of approximate orthogonality of Gaussian functions in a superposition appears physically appropriate, given the finite resolution of real-world measuring devices and the need to work with equivalence classes of states that are indistinguishable by the device.

Consider the space $V$ of finite linear combinations of Gaussian functions $g_c$ used earlier in the paper, where $c = z_k$ corresponds to some partition $\{z_k\}$, $ k=1,2,...,  N$ ($N \ge 3$) of the $z$-axis. Assume, as before, that the functions $g_c$ are sufficiently narrow so that the orthogonality condition holds for different $z_k$ and $z_m$ in the partition. In this setting, the proof of the orthogonality of $d\varphi_\tau/d\tau$ and $d\varphi_\lambda/d\lambda$ follows the same reasoning as presented earlier. Thus, regardless of the specific form of $\varphi$ in $V$, in particular, regardless of how many Gaussian functions appear in $\varphi$, these two directions at $\varphi$ remain orthogonal.

For an arbitrary state $\varphi \in V$, where $V$ is as defined above, consider the manifold $M_\varphi$ with coordinates $(\tau, \lambda)$, as defined in (\ref{surf}). (The fact that not all functions in $M_\varphi$ belong to $V$ does not pose a problem, since $M_\varphi$ lies in $L_2(\R)$, the space where orthogonality is required, and any function in $L_2(\R)$ can be approximated by a finite linear combination of the $g_c$'s from some space $V$.) For each point $(\tau, \lambda) \in M_\varphi$, consider the set of all functions in $V$ with $\mu_z = \tau$ and $\delta_z = \lambda$. This defines a foliation of $V$ of codimension 2. The leaves $\{\varphi\}_{\tau, \lambda}$ consist of all functions in $V$ that share the same values of $\mu_z$ and $\delta_z$. As discussed, the $\tau$ and $s=\ln \lambda$ coordinates on each $M_\varphi$ form Cartesian coordinates that identify $M_\varphi$ with $\R^2$. The corresponding components of a step of the random walk in {\bf (RM)} from any $\varphi$ are independent random variables.

Due to the homogeneity of the step distribution in {\bf (RM)}, the probability distributions for the $\tau$ and $s$ components are identical at all points $\varphi$. By definition, $\mu_z$ and $\delta_z$ do not vary along the leaves. Moreover, the values of $\mu_z$ and $\delta_z$ are the only quantities needed to determine whether the state has reached an equivalence class of the position eigenstate. Components of a step from $\varphi$ that are tangent to the leaf through $\varphi$ do not affect $\mu_z$ and $\delta_z$, and therefore do not contribute to collapse to a physical eigenstate of position. Thus, the walk on the manifold $M_\varphi$, identified with the $(\tau, s)$ plane $\R^2$, is sufficient to describe the collapse in this setting.

The set of points with $\delta_z < \delta$ on the $s$-axis forms a half-line. As the number of steps increases, the probability that the state under the random walk in {\bf (RM)} satisfies this condition approaches $1/2$. A drift in the $s$-coordinate on the leaf $\{\varphi\}_{\tau, \lambda}$ can guide the state toward the $z$-axis, increasing the probability of satisfying $\delta_z < \delta$ to nearly 1 after just a few steps, given a suitable choice of parameters. Meanwhile, the walk in $\tau$ determines the relative probabilities of reaching eigenstates, in accordance with the Born rule. A possible physical origin of the drift will be examined in Section IV.

 As established, the collapse process in the model reduces to a random walk of the initial state on the manifold $M_\varphi$, identified with the $(\tau,s)$ plane $\R^2$.
The $\tau$ and $s$ components of the steps in the random walk, generated by the Hamiltonian ${\widehat h}$ in {\bf (RM)}, are independent identically distributed normal random variables. 
The walk of the initial state along $M_\varphi$ consists of a random walk without drift in the $\tau$-coordinate and a random walk with drift in the positive direction of the $s$-axis. In other words, the walk is represented as follows:
\begin{equation}
\tau_k=\tau_{k-1}+\xi_k
\end{equation}
and
\begin{equation}
s_k=s_{k-1}+h + \eta_k,
\end{equation}
where $\xi_k$ and $\eta_k$ are independent identically distributed normal random variables, and $h$ is a positive number equal to the step of the drift. Using $s_0=0$, we have, for the $N$-th step of the walk  in $s$:
\begin{equation}
s_N=h\cdot N+\sum^{N}_{k=1} \eta_k.
\end{equation}
Given that $\lambda=e^s$ and $\delta_z=\lambda^{-1} \delta_{z_0}$, we see that $\delta_z=e^{-s}\delta_{z_0}$. Therefore, the variance exponentially approaches zero with an increase in $s$. In this case, even a few steps of the walk of the state may be sufficient to reach a neighborhood of the $z$-axis. The gambler's ruin process in the variable $\tau$ in the second example is then guaranteed to take the state to $\{g_a\}$ or $\{g_b\}$ with the probability satisfying the Born rule, as derived in (\ref{born}). The time interval of collapse (the mean observation period) of a given state $\varphi$ in the model depends on the frequency and the distribution of steps of the walk, the value of the parameter $h$, and the parameters in the definition of the equivalence classes $\{g_a\}$ and $\{g_b\}$. 

A computer simulation illustrating three runs of the random walk of the initial state $\varphi=\alpha g_a+ \beta g_b$ in the second example, where $|\alpha|=1/2$ and $|\beta|={\sqrt 3}/2$, is depicted in Figure 2. The simulation uses parameters $a=-10$, $b=10$, step sizes $|\xi_k|=|\eta_k|=1$, and a drift parameter $h=1/2$. The figure illustrates the evolution of the expected value $\tau_n$ and the standard deviation $\delta_n$ of the $z$-coordinate throughout the walk.
Figure 3 displays a bar graph showing the counts of the state reaching the detector at positions $a$ and $b$ over 1500 runs of the walk, using the same parameters, along with its consistency with the Born rule.
\begin{figure}[ht]
\centering
\includegraphics[width=8cm]{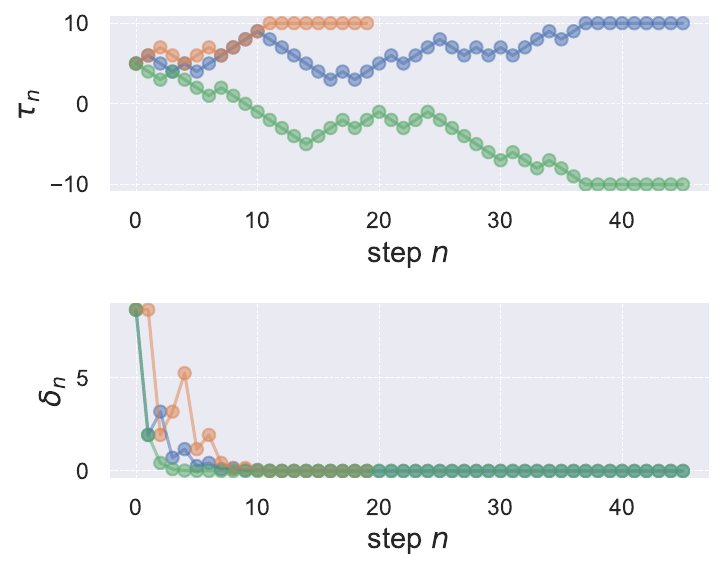}
\caption{Computer simulation of three runs of the random walk.}
\label{fig:2}
\end{figure}

\begin{figure}[ht]
\centering
\includegraphics[width=8cm]{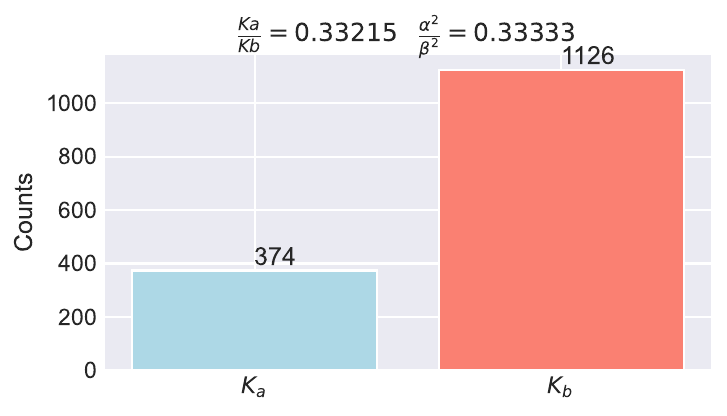}
\caption{Relative frequency of reaching the detector.}
\label{fig:3}
\end{figure}

If the drift parameter $h$ vanishes, on average, more steps will be needed to satisfy both conditions, $\delta_z<\delta$ and $\mu_z=a$ or $\mu_z=b$. Nevertheless, if $s$ is unbounded in $\R$, the probability of $s_N$ meeting the half-line condition $\delta_z<\delta$ will converge to $1/2$ as the number of steps increases; in the bounded case, this probability will be even higher. Given that only a fraction of particles in a double-slit experiment reach the detector, this result indicates that the proposed collapse model, based on {\bf (RM)} and equivalence classes of eigenstates, remains feasible even in the absence of drift. 
A detailed analysis of the model with various values of the parameters will be presented in a separate work.

There is an interesting geometric interpretation that relates the considered walk on $M_\varphi$ with a walk of a spin-state $[\alpha, \beta]$ on the sphere $S^2=\mathbb{CP}^1$.
Namely, by a proper choice of the unit and the origin on the $z$-axis, one can always ensure that $a=-1$ and $b=1$. 
With this, we have for the initial state $\varphi=\alpha g_a+\beta g_b$:
\begin{equation}
\label{z1}
\mu_z=|\beta|^2-|\alpha|^2
\end{equation}
and
\begin{equation}
\label{sig}
\delta^2_z=1-\mu^2_z=4|\alpha|^2|\beta|^2.
\end{equation}
Expressions (\ref{z1}) and (\ref{sig}) are intimately related to the expressions for Cartesian coordinates of the spin-state $[\alpha, \beta] \in \C^2$ under the usual bundle projection $\pi: S^3 \longrightarrow \mathbb{CP}^1=S^2$. These coordinates are given by 
\begin{eqnarray}
\label{xyz}
x&=&\alpha{\overline \beta}+{\overline \alpha}\beta,\\
\label{xyz1}
y&=&i(\alpha{\overline \beta}-{\overline \alpha}\beta), \\
\label{xyz2}
z&=&|\beta|^2-|\alpha|^2.
\end{eqnarray}
From these equations, we see that $\mu_z=z$ and $\delta^2_z=x^2+y^2$.  The coefficients $\alpha$ and $\beta$ of $\varphi$ may also have a  phase difference $\theta$. Adding the variable $\theta$ to the pair $(\mu_z, \delta_z)$, we obtain cylindrical coordinates on the sphere $S^2$. 

For our third realization of the model, let us use the triple $(\mu_z, \delta_z, \theta)$ to describe the walk of state $\varphi$ as a motion on the sphere. Namely, given a state $\psi=\alpha \widetilde{g}_{c}+ \beta \widetilde{g}_{d}$ evolving on $M_\varphi$, we identify the coordinates $(\mu_z, \delta_z, \theta)$ and find the corresponding point $(x,y,z)$ on the sphere with the help of equations (\ref{z1}-\ref{xyz2}).  In such a way, we identify the change in the values of $c$ and $d$ with the corresponding change in the coefficients $\alpha$ and $\beta$ of the initial state $\varphi=\alpha g_a+ \beta g_b$. In this case, the basis states $g_a$ and $g_b$ remain fixed during the evolution while the values of the coefficients $\alpha$ and $\beta$ are obtained from the equations  (\ref{z1},\ref{sig}). The evolution of the state is thus confined to the space $\C^2$.

The issue with this geometric realization of the evolution is that it imposes a relationship between $\mu_z$ and $\delta_z$. Specifically, it necessitates $\delta^2_z = 1 - \mu^2_z$, which is invalid when the state, during its evolution, is not confined to the space $\C^2$.
Note also that the change in the parameter $\theta$ during the walk cannot be determined from equations (\ref{z1}) and (\ref{sig}).
However, imposing the relationship $\delta^2_z=1-\mu^2_z$ without changing the walk in $\tau$ preserves the probabilities of reaching the eigenstates. Furthermore, it makes reaching the values $\mu_z=a$ of $\mu_z=b$ equivalent to reaching the eigenstates, which is similar to what the drift in $s$ has achieved. Although imposing this relationship is rather arbitrary, 
the change in $\mu_z=z$ and $\delta^2_z=x^2+y^2$ when the state approaches the poles of $S^2$ gives us a nice illustration of collapse in the model.
Note that the actual random walk of state studied in the paper does not happen on $\mathbb{CP}^1=S^2$, which, unlike the manifold $M_\varphi$, does not even include the $z$-axis. In particular, the walk does not converge to Brownian motion on the sphere.

\section{Directions for future research}

Thus far, the presented approach to {\bf (RM)} and the drift has been primarily mathematical. 
Several immediate tasks must be addressed to further develop the proposed model. First, it is highly desirable to present clear arguments supporting the use of random matrices and the conjecture {\bf (RM)} to address the problem of quantum measurement and the transition to classicality. The results should also be compared with those of well-developed theories of decoherence and quantum measurement, including models of continuous measurement \cite{Decoh, J, SJ}.
Second, it is important to investigate how the parameters of the random walk in {\bf (RM)} relate to the properties of the measuring device, the environment, and the system itself, and how these properties influence the diffusion coefficient ${\mathbb D}$.
Third, in the version of the model that includes a drift term, it is necessary to establish a plausible physical origin for the drift.
In Section VI, we take initial steps toward addressing the first task, while our focus here is on the third, leaving the second for a forthcoming paper.

As previously discussed, the drift term, responsible for the squeezing of the state function, is important for explaining the stability of measurement outcomes and for understanding the dynamics of macroscopic bodies in the context of Schr{\"o}dinger dynamics constrained to $M^{\sigma}_{3,3}$. 
Furthermore, under the assumption of {\bf (RM)} and the macroscopicity condition ${\mathbb D}\approx 0$ (which eliminates the stochasticity of evolution during measurement), the process responsible for the drift must be nearly deterministic. That is, like the spreading under conventional Schr{\"o}dinger evolution, the squeezing must occur with near certainty.
A natural example of the state-squeezing phenomenon, which may be associated with drift in the parameter $s$, is observed in spontaneous emission. When an atom transitions from a higher to a lower energy state and emits a photon, the electron's state function narrows, converging to the spatial extent of the atom in its ground state.
Under typical conditions, the process is nearly deterministic in the sense that the transition to a lower energy level occurs with near certainty.
Although explaining such transitions requires the use of a quantized electromagnetic field, the theory has been well established since Dirac's seminal work \cite{Dirac}. The governing equations are linear; however, the process becomes non-unitary and irreversible due to the averaging over photon modes required to compute the transition amplitude.

 A similar process that causes the narrowing of a particle's state function occurs spontaneously and universally under general conditions in molecules or when a particle is confined to a 
 potential well. The released energy may manifest as radiation or phonons. 
 It is proposed that under proper conditions this process may be responsible for the narrowing of the state function of a measured particle during its interaction with a measuring device, such as a scintillation screen. 
In practice, this interaction with the screen involves exciting numerous atoms, resulting in spontaneous emission until the particle, assumed to be distinct from the material's particles, loses most of its energy and becomes trapped by one of the molecules.

When the potential of the system, consisting of the weakened particle and particles of the screen it interacts with, can be assumed harmonic, the energy levels of the trapped particle are given in natural units by $E_n = \frac{1}{2} + n$. In this scenario, a calculation yields the expression for the variance as $\delta^2_{n} = \frac{1}{2} + 4n$. As the excited state descends the ladder of energy levels, the standard deviation for the state decreases to a small value, comparable to the size of a molecule of the screen.
The same occurs for a potential that is approximately harmonic or quartic near the stable point, or, more generally, is U-shaped. 
While the particle may not reach the ground state in practice, and the ground state itself may not be a Gaussian state, these specifics are inconsequential. What matters is that the state function ``contracts" to a sufficiently small size to fall within the equivalence class of a position eigenstate.

The physical basis for this drift under measurement remains speculative and will be explored in future work. The idea is to construct a full Hamiltonian incorporating: (1) the standard Schr{\"o}dinger Hamiltonian (with potential, if present), (2) the {\bf (RM)} Hamiltonian, (3) the electromagnetic field Hamiltonian, and (4) the interaction Hamiltonian between the particle and the field. The relative significance of these terms depends on how the particle's position is measured.
Consider the following three scenarios.

\textbf{The position of a microscopic particle is measured by scattering of light.} The {\bf (RM)} term dominates, while other components of the Hamiltonian are negligible over the observation period. The Born rule emerges as the state localizes near one of the slits (in the double-slit experiment). After localization, the particle may shift and be detected elsewhere (i.e., it is not trapped). Photons involved are assumed to be sufficiently weak.
    
\textbf{The position of a microscopic particle is measured by a scintillation screen.} Both the {\bf (RM)} term and the ``spontaneous emission term" (the interaction Hamiltonian associated with a process similar to spontaneous emission resulting in a drift toward $M^{\sigma}_{3,3}$) are relevant. The outcome is consistent with the Born rule and additionally involves localization to $M^{\sigma}_3$ and trapping within the potential of a screen atom or molecule.
    
 \textbf{A macroscopic particle whose postion is measured by the environment.} The {\bf (RM)} term becomes negligible (${\mathbb D} \approx 0$). Schr{\"o}dinger evolution and ``spontaneous emission" combine, with the latter inducing drift toward $M^{\sigma}_{3,3}$, ultimately recovering Newtonian dynamics.

To reiterate, these scenarios are only outlined briefly here and are not explored in the present paper. This work focuses on two scenarios: measurement governed by {\bf (RM)} without drift, and measurement governed by {\bf (RM)} with a mathematical drift in standard deviation (i.e., toward $M^{\sigma}_{3}$). In both scenarios, measured eigenstates are defined via equivalence classes.

\section{Electrons versus bullets}

In a version of Feynman's experiment with bullets, a machine gun shoots a stream of bullets into a screen with two slits. Behind the slits, there is a wooden screen that absorbs bullets. A small movable sandbox in front of the screen is used as a detector of bullets along the $z$-axis on the screen. 
The setup of this experiment is, therefore, very similar to the one with a microscopic particle such as electron considered in the paper. 
Furthermore, we saw that classical space $\R^3$ is isometric to the submanifold $M^{\sigma}_{3}$ of the space of states $\mathbb{CP}^{L_2}$. A point ${\bf a}$ in classical space $\R^3$ is represented by the state $g_{{\bf a},\sigma}$ in $M^{\sigma}_{3}$, defined in (\ref{g}).
Similarly, the classical phase space $\R^3\times \R^3$ for a particle is isometric to the submanifold $M^{\sigma}_{3,3}$ of the space of states of the particle. Most importantly, it was verified that Newtonian motion of a particle is equivalent to the Schr{\"o}dinger evolution of its state, provided the state is constrained to the manifold $M^{\sigma}_{3,3}$.
Based on that, we can identify the path of a classical particle with the corresponding path in $M^{\sigma}_{3,3}$  in a physically meaningful way.  
In particular, neglecting other coordinates in $\R^3$, the path $z=c(t)$ of a particle going through point $a$ in $\R$ is represented by the path $\varphi=g_{c(t)}$ of its state going through the point $g_a$ in $L_{2}(\R)$.
This mathematically rigorous and physically valid identification, together with the conjecture {\bf (RM)}, give us a perfect setup for analyzing and comparing the double-slit experiments with electrons and bullets.

Let us consider the experiment with electrons first. The electron's spin properties in the experiment will be neglected. 
At the beginning of the experiment, an electron gun fires electrons one by one.
We may assume that state of the initial electron is a Gaussian wave packet moving towards the screen with the slits. In particular, the state is near the manifold $M^{\sigma}_{3,3}$ in the space of states $\mathbb{CP}^{L_2}$. That is, the Fubini-Study distance from the state to $M^{\sigma}_{3,3}$ is small. During this time, the state propagates by the usual Hamiltonian ${\widehat h}=\frac{{\widehat {\bf p}^2}}{2m}+{\widehat V}({\bf x})$, where ${\widehat V}({\bf x})$ is an external potential including the one associated with the screen with the slits. Interaction of the electron with the surrounding matter in the experiment can be neglected. Upon interaction with the screen, the wave packet splits into a superposition of two wave packets. That means that the state is no longer on the manifold $M^{\sigma}_{3,3}$. In fact, assuming, for example, that $\varphi=\alpha g_a+\beta g_b$ with  $|\alpha| \le |\beta|$, the cosine of the smallest distance between the state and $M^{\sigma}_{3,3}$ is given by 
\begin{equation}
|(\alpha g_a+\beta g_b, g_b)|=|\beta|.
\end{equation}
It follows that the state is close to $M^{\sigma}_{3,3}$ only when $\alpha$ is close to $0$. This is not the case immediately to the right of the screen with both slits open.

Note that nothing special has happened to the state at this time. It simply moved away from the classical phase space submanifold $M^{\sigma}_{3,3}$ into $\mathbb{CP}^{L_2}$. In particular, the path of the state did not go through the points $g_a$ or $g_b$, or any other point $g_c$ with $c$ on the $z$-axis. It passed in the space of states ``over" the $z$-axis and the screen. However, for the electron to have any position  in $\R^3$ at all, the electron's state must be in $M^{\sigma}_{3,3}$, which is not the case when the electron interacts with the screen. So, the electron position is not defined at this time. It is not given by $a$ or $b$ on the $z$-axis, or by any other point in $\R^3$. At the same time, whenever the electron's state {\it is} in $M^{\sigma}_{3,3}$, it identifies the electron's position in $\R^3$ correctly, as a dynamical variable, in a way consistent with Newtonian dynamics. In this sense, the state variable $\varphi$ is an {\it extension} of the classical position variable of the particle. Instead of saying that the electron's position is not defined when the particle interacts with the screen, we can say that the electron's path takes off the classical space and passes ``over" the screen in the space of states. Its position along the path is well-defined but requires additional dimensions provided by the space of states $\mathbb{CP}^{L_2}$. In particular, the electron's path does not ``split" to go through two slits at once.
It is only when we insist that the electron's state must always be on $M^{\sigma}_{3,3}$ that we run into this paradox.

What happens to the right of the screen, when the particle interacts with the detector? The Born rule for the probability density function for the particle's position, in the considered approximation, yields  $P(z)=|\alpha g_a(z)+\beta g_b(z)|^2=|\alpha|^2|g_a(z)|^2+|\beta|^2 |g_b(z)|^2$. Integrating this over the area occupied by the detector near point $a$, we get approximately $|\alpha|^2$. The probability of being near $b$ is then $|\beta|^2$. 
This result is identical to the one obtained from the conjecture {\bf (RM)} in the paper.
According to {\bf (RM)}, the state $\varphi$ is driven by the Hamiltonian represented by a random matrix. The random walk of state brings it back to the classical space submanifold $M^{\sigma}_{3}$ to the equivalence class of one of the eigenstates $g_a$ or $g_b$ by the process described in the previous sections. 
The electron is then positioned near the point $a$ or point $b$  with the probabilities $|\alpha|^2$ and $|\beta|^2$ respectively.

Suppose now that the detected particle is able to continue its motion towards the screen on the right of the detector. It will then arrive at the screen as a spread-out version ${\widetilde g}_a$ (or ${\widetilde g}_b$) of the detected Gaussian state $g_a$ (or $g_b$). The probability density function for the electron's position on the screen is then given by either $P(z)=|{\widetilde g}_a(z)|^2$ or $|{\widetilde g}_b(z)|^2$ and no interference pattern is observed on the screen. The resulting ``corpuscular" properties of the detected electron are due to the closeness of its ``post-detector" state to the classical phase space manifold $M^{\sigma}_{3,3}$ during its motion from the detector to the backstop screen. As we know, when the electron's state is {\it on} $M^{\sigma}_{3,3}$, it satisfies Newtonian dynamics and behaves like a particle.

 If the experiment is repeated without the detector, the state $\varphi=\alpha g_a +\beta g_b$ obtained to the right of the slits will continue its motion towards the backstop screen along a path that is away from $M^{\sigma}_{3,3}$. Interaction of the particle with the backstop screen happens in the same way as its interaction with the detector. However, this time the spread-out states ${\widetilde g}_a$ and ${\widetilde g}_b$ may not be considered orthogonal. As shown earlier in this paper and in \cite{KryukovNew}, the conjecture {\bf (RM)}, when applied to this case, yields the Born rule as before. 
 Provided the particle has been detected by the screen, the probability density function for the position is given by $P(z)=|\alpha{\widetilde g}_a(z)+\beta {\widetilde g}_b(z)|^2$. The interference term is now present. 
 The observed ``wave" properties of the electron are caused by its state being distant from the classical phase space submanifold $M^{\sigma}_{3,3}$ during its motion from the screen with the slits to the backstop screen.  That is, the state arrives at the backstop screen as a superposition $\alpha {\widetilde g}_a +\beta {\widetilde g}_b$, and such a superposition is away from $M^{\sigma}_{3,3}$.
When the state of the particle in the experiment moves away from the classical phase space submanifold  
$M^{\sigma}_{3,3}$, the standard deviation $\delta_z$ increases and the particle demonstrates its wave properties. When the state is brought back to the manifold $M^{\sigma}_{3,3}$, the standard deviation decreases, and the particle demonstrates classical corpuscular properties.

What is different about the experiment with bullets? 
Measuring the position of a small electron in the experiment requires a detector or a backstop screen that the electron interacts with. On the other hand, the bullet interacts randomly and continuously in time with particles of the surroundings even before it reaches the sandbox or the backstop screen. 
Because of this continuous interaction, the surroundings (particles of air, radiation) contain information about the bullet's position at all times. In other words, the bullet's position is constantly measured by the surroundings. 
It follows that the conjecture {\bf (RM)}, when accepted, needs to be applied to the entire motion of the bullet in the experiment.

As shown in Section II, the state driven by the Hamiltonian in {\bf (RM)}, and conditioned to stay on the manifold $M^{\sigma}_{3}$, describes the Brownian motion of the particle. When the particle is sufficiently large, the diffusion coefficient for the Brownian motion vanishes, and the particle is at rest in the lab system. The isotropy of the probability distribution of steps of the random walk of the state signifies that the state of the particle in the space of states $\mathbb{CP}^{L_2}$ must then be at rest as well. If an external potential is applied to such a system, the particle, under the accepted assumptions (i.e., {\bf (RM)} and $M^{\sigma}_{3,3}$ as the classical phase space, or, alternatively, {\bf (RM)} combined with the drift), will move in accordance with Newtonian dynamics. A bullet is large enough for its Brownian motion in natural environment to be trivial. 
It follows that the state of the bullet is confined to $M^{\sigma}_{3,3}$. Thus, the dynamics of the bullet in the framework is described by Newton's equations of motion.

\section{Why random matrices?}

The conjecture {\bf (RM)} provides a unified model of measurement that applies to both macroscopic and microscopic particles. The constraint relating the random walk on the state space to the corresponding random walk on the manifold $M^{\sigma}_3$
is consistent with the condition established in Section II, which connects Schr{\"o}dinger and Newtonian dynamics. Since Brownian motion can, under certain statistical assumptions, be derived from the Newtonian dynamics of a particle in a thermal bath, and since the distribution of steps in the random walk on 
$M^{\sigma}_3$ defines the Gaussian unitary ensemble in {\bf (RM)}, this opens the possibility of supporting {\bf (RM)} through underlying dynamical considerations.
Moreover, the translational and rotational symmetries observed in macroscopic measurements are preserved in the model. The irreversibility of measurement arises from the lack of time-reversal invariance in Hamiltonians drawn from the Gaussian unitary ensemble \cite{KryukovNew}, and potentially also from the inherent irreversibility of the process generating the drift. The model not only leads to a derivation of the Born rule but also accounts for the outcomes of the double-slit experiment, both with and without a detector.
These compelling results and the new avenues of research they open provide indirect support for the conjecture. However, a fundamental question remains: why should the Hamiltonian during measurement be represented by a random matrix?

Random matrices were introduced into quantum mechanics by Wigner \cite{Wigner} in a study of excitation spectra of heavy nuclei. 
Wigner reasoned that the complexity of the motion of nucleons in the nucleus could be handled by modeling the Hamiltonian of the system with a random matrix. The ensemble of matrices only had to respect the symmetries of the system. The correlations in the spectrum of random matrices that Wigner discovered turned out to be applicable to a remarkably large number of quantum systems with many as well as few degrees of freedom. Experimental evidence suggests that all quantum systems whose classical counterpart is chaotic demonstrate random matrix statistics, as proposed in the Bohigas-Giannoni-Schmit (BGS) conjecture  \cite{BGS}. On another note, classical measurement can be modeled by Brownian motion. It is known that Brownian motion can be characterized as a chaotic process \cite{Nature1, Nature2,  Cecconi}. The complex nature of the interaction between the measured particle and atoms of the detector, coupled with the chaotic features of Brownian motion, suggests that the system's Hamiltonian can be effectively represented by a random matrix.

Decoherence theory \cite{Decoh} seeks to explain the process of position measurement based on the Schr{\"o}dinger evolution of the system interacting with the environment. A typical Hamiltonian modeling this situation would describe a particle linearly coupled to a set of harmonic oscillators. Alternatively, the scattering matrix can be used to determine the effect of the collective scattering of particles on the particle whose position is measured. The evolution of the density matrix of the measured particle would then exhibit a damping of interference terms in the matrix. The theory has been successful in explaining the emergence of classical probabilities. However, it falls short in explaining how a single classical outcome arises as a result of measurement \cite{AdlerWhy} and does not lead to the Born rule. 

In loose terms, deriving evolution equations for the density matrix in decoherence theory is akin to attempting to derive Brownian motion from the Newtonian dynamics of a system of particles. While both endeavors offer proof of concept, they rely on several crucial assumptions and fall short of providing a fundamental explanation of the phenomena. 
For example, deriving Brownian motion typically involves making simplifying assumptions about the form of the Hamiltonian (such as a harmonic bath and bilinear interaction) and the spectral density. Attempting to derive Brownian motion as the limit of a deterministic system of hard spheres is mathematically highly complex and also requires additional assumptions beyond Newtonian dynamics \cite{HardSphere}.
Ultimately, these endeavors serve as useful models. However, to significantly simplify the description and gain deeper insight into the phenomena, additional symmetry-based assumptions about the dynamics, such as those proposed by Einstein in the theory of Brownian motion or by Wigner in the study of spectra of heavy nuclei, are still necessary.
Similarly, the universal applicability of random matrix theory to fluctuations in quantum systems, together with the results derived here, suggests that random matrices may offer a simplifying mechanism and the missing insight into the process of measurement.



\section*{} 

\end{document}